\newif\ifpdf
\ifx\pdfoutput\undefined
\pdffalse 
\else
\pdfoutput=1 
\pdftrue \fi

\newif\ifarxiv
\arxivtrue

\documentclass[cmp,final]{svjour}
\usepackage{cite}
\usepackage{amsmath}
\usepackage{amsfonts}
\usepackage{amssymb}
\usepackage{epsf}
\usepackage{graphicx}
\usepackage{graphics}
\usepackage{verbatim}
\usepackage{epsfig}

\IfFileExists{myowntimes.sty}
	{\usepackage{myowntimes}}
	{\usepackage{times}\usepackage{mathrsfs}}
	
\DeclareFontFamily{OT1}{eusm}{} \DeclareFontShape{OT1}{eusm}{m}{n} {<5> <6> <7> <8> <9> <10> <11> <12> <14.4> eusm10}{}
\DeclareMathAlphabet{\eusm}{OT1}{eusm}{m}{n}

\DeclareFontFamily{OT1}{fraktura}{}
\DeclareFontShape{OT1}{fraktura}{m}{n} {<5> <6> <7> <8> <9> <10> <11> <12> <13> <14.4> [1.1] eufm10}{}
\DeclareMathAlphabet{\fraktura}{OT1}{fraktura}{m}{n}

\ifarxiv
\fi

\newenvironment{proofsect}[1]{\vskip0.1cm\noindent{\rmfamily\itshape #1.}}{\qed\vspace{0.15cm}}

\spnewtheorem{mylemma}[theorem]{Lemma}{\bf}{\it}
\spnewtheorem{myproposition}[theorem]{Proposition}{\bf}{\it}
\spnewtheorem{mycorollary}[theorem]{Corollary}{\bf}{\it}
\spnewtheorem{mydefinition}[theorem]{Definition}{\bf}{\it}
\spnewtheorem{myremark}[theorem]{Remark}{\it}{\rm}
\spnewtheorem{myremarks}[theorem]{Remarks}{\it}{\rm}
\numberwithin{equation}{section}
\numberwithin{theorem}{section}

\newcommand{\dist}{\operatorname{dist}}

\newcommand{\textd}{\text{\rm d}\mkern0.5mu}
\newcommand{\texti}{\text{\rm  i}\mkern0.7mu}
\newcommand{\texte}{\text{\rm e}}

\newcommand{\1}{\text{\sf 1}}

\renewcommand{\AA}{\mathcal A}
\newcommand{\BB}{\mathcal B}
\newcommand{\CC}{\mathcal C}

\newcommand{\GG}{\mathcal G}
\newcommand{\HH}{\mathcal H}
\newcommand{\II}{\mathcal I}

\newcommand{\LL}{\mathcal L}

\newcommand{\CalS}{\mathcal S}

\newcommand{\C}{\mathbb C}

\newcommand{\BbbP}{\mathbb P}

\newcommand{\R}{\mathbb R}

\newcommand{\T}{\mathbb T}

\newcommand{\Y}{\mathbb Y}
\newcommand{\Z}{\mathbb Z}

\newcommand{\scrC}{\mathscr{C}}

\newcommand{\scrI}{\mathscr{I}}
\newcommand{\scrJ}{\mathscr{J}}

\newcommand{\scrM}{\mathscr{M}}

\newcommand{\scrS}{\mathscr{S}}

\newcommand{\twoeqref}[2]{(\ref{#1}--\ref{#2})}
\newcommand{\cc}{{\text{\rm c}}}

\def\myffrac#1#2 in #3{\raise 2.6pt\hbox{$#3 #1$}\mkern-1.5mu\raise 0.8pt\hbox{$#3/$}\mkern-1.1mu\lower 1.5pt\hbox{$#3 #2$}}
\newcommand{\ffrac}[2]{\mathchoice%
	{\myffrac{#1}{#2} in \scriptstyle}
	{\myffrac{#1}{#2} in \scriptstyle}
	{\myffrac{#1}{#2} in \scriptscriptstyle}
	{\myffrac{#1}{#2} in \scriptscriptstyle}
}

\newcommand{\done}[2]{\sqrt{\!\CalS}\,\|#1-#2\|_1}
\newcommand{\dtwo}[2]{\,\CalS\|#1-#2\|_2^2}
\newcommand{\dall}[2]{\,\textd_{\CalS}(#1,#2)}
\newcommand{\norm}[2]{\Vert #1-#2\Vert}
\newcommand{\bra}[1]{\langle#1|}
\newcommand{\ket}[1]{|#1\rangle}
\newcommand{\ip}[2]{\langle{#1\vert#2}\rangle}
\newcommand{\floor}[1]{\lfloor{#1}\rfloor}
\newcommand{\ceil}[1]{\lceil{#1}\rceil}
\newcommand{\TR}{\text{\rm Tr}}
\newcommand{\state}[1]{\left\langle #1\right\rangle}
\newcommand{\proj}[1]{\ket{#1}\bra{#1}}

\newcommand{\wt}{\widetilde}
\newcommand{\whvartheta}{\hat\vartheta}

\newcommand{\wh}{\widehat}
\newcommand{\br}{\boldsymbol r\mkern1.5mu}
\newcommand{\bk}{\boldsymbol k\mkern1.5mu}
\newcommand{\bS}{\boldsymbol S}

\newcommand{\frakA}{\fraktura A}
\newcommand{\bt}{\boldsymbol t}
\newcommand{\frakp}{\fraktura p}
\newcommand{\frakq}{\fraktura q}
\newcommand{\hatQ}{\hat Q}
\newcommand{\betat}{\beta_{\text{\rm t}}}
\newcommand{\hate}{\hat{\text{\rm e}}}
\newcommand{\hatv}{\hat{\text{\rm v}}}
\newcommand{\hatw}{\hat{\text{\rm w}}}
\newcommand{\BBE}{\BB_{\text{\rm E}}}
\newcommand{\BBSW}{\BB_{\text{\rm SW}}}

\newcommand{\eusmP}{\eusm P}
\newcommand{\frakc}{\fraktura c}
\newcommand{\frakE}{\fraktura E}
\newcommand{\frake}{\fraktura e}

\newcommand{\mix}{\text{\rm mix}}

\newcommand{\ord}{\text{\rm ord}}
\newcommand{\dis}{\text{\rm dis}}
\newcommand{\Gso}{\GG_{\text{\rm ord}}}
\newcommand{\Gsd}{\GG_{\text{\rm dis}}}  

\begin{document}

\title{Quantum spin systems at positive temperature}
\titlerunning{Quantum spin systems at positive temperature}
\author{Marek Biskup\and Lincoln Chayes\and Shannon Starr}
\authorrunning{M.~Biskup, L.~Chayes and S.~Starr}
\institute{Department of Mathematics, UCLA, Los Angeles, CA 90095, U.S.A.}
\date{}
\maketitle

\renewcommand{\thefootnote}{}
\footnotetext{\vglue-0.41cm\footnotesize\copyright\,2006 by M.~Biskup, L.~Chayes and S.~Starr. Reproduction, by any means, of the entire article for non-commercial purposes is permitted without charge.}
\renewcommand{\thefootnote}{\arabic{footnote}}

\begin{abstract}
We develop a novel approach to phase transitions in quantum spin models based on a relation to their classical counterparts. Explicitly, we show that whenever chessboard estimates can be used to prove a phase transition in the classical model, the corresponding quantum model will have a similar phase transition, provided the inverse temperature~$\beta$ and the magnitude of the quantum spins~$\CalS$ satisfy~$\beta\ll\sqrt\CalS$. From the quantum system we require that it is reflection positive and that it has a meaningful classical limit; the core technical estimate may be described as an extension of the Berezin-Lieb inequalities down to the level of matrix elements. The general theory is applied to prove phase transitions in various quantum spin systems with $\CalS\gg1$. The most notable examples are the quantum orbital-compass model~on~$\Z^2$ and the quantum 120-degree model on~$\Z^3$ which are shown to exhibit symmetry breaking at low-temperatures despite the infinite degeneracy of their (classical) ground state.
\end{abstract}

\setcounter{tocdepth}{3}
\footnotesize
\tableofcontents
\normalsize

\section{Introduction}
\label{sec1}\noindent
It is considered common knowledge that, for spin systems, the behavior of a quantum model at finite temperature is ``like'' the behavior of the corresponding classical model.  However, beyond the level of heuristics, it is far from clear in what sense the above statement is meaningful. Another, slightly more academic way to ``recover'' the classical spin system is to consider spin-representations with spin-magnitude~$\CalS$ and then let~$\CalS\to\infty$. A standard argument as to why this should work is that the commutators between various spin operators are order-$1/\CalS$ smaller than the quantities themselves, and so the spins behave essentially classically when~$\CalS$ is large. Notwithstanding, precise statements along these lines have only been made for the $\CalS\to\infty$ limit of the free energies~\cite{Berezin,Lieb,Fuller-Lenard1,Fuller-Lenard2,Simon} and specific types of~$1/\CalS$ corrections~\cite{Conlon-Solovej,Michoel-Nachtergaele1,Michoel-Nachtergaele2}.

A common shortcoming of the above studies is that neither spells explicit conditions on the relative magnitude of~$\beta$ and~$\CalS$ for which the classical behavior is exhibited. This is of importance because, at sufficiently low temperatures, the relevant excitations are \emph{quantum}. For example, while the classical Heisenberg antiferromagnet on a finite bipartite graph has a continuum of ground states (related by the SO(3) symmetry), the quantum Heisenberg antiferromagnet has a unique ground state~\cite{Lieb-Mattis}. Another example is the 111-interface in the classical Ising model which, at zero temperature, is disordered but may be stabilized by appropriate (but arbitrarily small) quantum perturbations~\cite{Koma-Bruno,Bolina-Shannon}. The control of the relevant quantum excitations is a non-trivial subject and is usually accomplished only when finite-temperature effects are of little significance for the overall~behavior. 

The preceding discussion is particularly important for systems which undergo phase transitions. Here several techniques have been available---infrared bounds \cite{DLS,FSS}, chessboard estimates \cite{FL,FILS1,FILS2,Kotecky-Shlosman} and contour expansions~\cite{BKU,DFF1,DFF2,KU}---some of which (specifically, the latter two) are more or less based on the assumption that the quantum system of interest has a strong classical component. However, while certain conclusions happen to apply uniformly well even as~$\CalS\to\infty$, the classical reference state of these techniques is usually \emph{discrete} (e.g., Ising type). This is quite unlike the~$\CalS\to\infty$ limit which inherently leads to a \emph{continuous-spin}, Heisenberg-like model. Thus, the relation between the above ``near-classical'' techniques and the $\CalS\to\infty$ results discussed in the first paragraph is tenuous.

The purpose of this paper is to provide a direct connection between the $\CalS\to\infty$ approach to the classical limit of quantum spin systems and the proofs of phase transitions by the traditional means of chessboard estimates.  Explicitly, we establish the following general fact: \textsl{Whenever chessboard estimates can be used to prove a phase transition in the classical system, a corresponding transition will occur in the quantum system provided~$\sqrt\CalS$ is sufficiently larger than the inverse temperature}. This permits us to prove phase transitions in systems with highly degenerate ground states, but without continuous symmetry, as well as certain temperature driven phase transitions which have not been accessible heretofore.

To highlight the main idea of our approach, let us recall how chessboard estimates enter the proofs of phase transitions. Suppose a quantum system on the torus is partitioned into disjoint blocks and a projector on a ``bad event'' is applied in some of the blocks. The goal is to show that the expectation---in the quantum Gibbs state---of the product of these projectors decays exponentially with the number of bad blocks. Here the chessboard estimates offer a non-trivial simplification: The expectation to the inverse number of bad blocks is maximized by the configuration in which all blocks are bad. In classical models, the latter quantity---sometimes referred to as the \emph{universal contour}---is often fairly easy to estimate by properly accounting for energy and entropy of the allowed configurations. However, this is not the case once quantum effects get into play; the only general technique that has been developed for this purpose is the ``principle of exponential localization''~\cite{FL} which hinges on an approximate diagonalization of the ``universal projectors'' and model-specific spectral estimates.

The main feature of our approach is that we bound the (relevant) universal contours directly---namely, by the universal contours for the \emph{classical} (i.e.,~$\CalS=\infty$) version of the quantum system. 
The technical estimate making this possible is a new bound on the matrix element of the Gibbs-Boltzmann weight relative to \emph{coherent states} $\ket\Omega$, which is close in the spirit to the celebrated Berezin-Lieb inequalities~\cite{Berezin,Lieb}. The result is that $\bra\Omega\texte^{-\beta H}\ket\Omega$ is dominated by the classical Gibbs-Boltzmann weight times a correction that is exponential in~$O(\beta/\sqrt{\CalS})\times\,$volume. Hence, if $\beta\ll\sqrt\CalS$, the exponential growth-rate of partition functions, even those constrained by various projectors, is close to that of the classical system. This is ideally suited for an application of chessboard estimates and the corresponding technology---developed in~\cite{FL,FILS1,FILS2,Kotecky-Shlosman}---for proving first-order phase transitions.
Unfortunately, the bound in terms of universal contour has to be performed before the ``conversion'' to the classical setting and so we still require that the quantum system is reflection positive.

To showcase our approach, we provide proofs of phase transitions in the following five quantum systems (defined by their respective formal Hamiltonians):
\settowidth{\leftmargini}{(11)}
\begin{enumerate}
\item[(1)]
The anisotropic Heisenberg antiferromagnet: 
\begin{equation}
\label{H1}
H=+\sum_{\langle\br,\br'\rangle}\CalS^{-2}(J_1 S^x_{\br}S^x_{\br'}+J_2S^y_{\br}S^y_{\br'}+S^z_{\br}S^z_{\br'})
\end{equation}
where~$0\le J_1,J_2<1$.
\item[(2)]
The non-linear XY-model:
\begin{equation}
\label{H2}
H=-\sum_{\langle\br,\br'\rangle}\eusmP\biggl(\frac{S^x_{\br}S^x_{\br'}+S^y_{\br}S^y_{\br'}}{\CalS^2}\biggr)
\end{equation}
where~$\eusmP(x)=\eusmP_1(x^2)\pm x\eusmP_2(x^2)$ for two polynomials~$\eusmP_1,\eusmP_2$ (of sufficiently high degree) with positive coefficients.
\item[(3)]
The non-linear nematic model:
\begin{equation}
\label{H3}
H=-\sum_{\langle\br,\br'\rangle}\eusmP\bigl(\CalS^{-2}(\bS_{\br}\cdot\bS_{\br'})^2\bigr)
\end{equation}
where $\bS_{\br}\cdot\bS_{\br'}=S^x_{\br}S^x_{\br'}+S^y_{\br}S^y_{\br'}+S^z_{\br}S^z_{\br'}$ and where~$\eusmP$ is a polynomial---typically of high degree---with positive coefficients.
\item[(4)]
The orbital compass model on~$\Z^2$:
\begin{equation}
\label{H4}
H=\sum_{\langle\br,\br'\rangle}
\begin{cases}
\CalS^{-2}\,S^x_{\br}S^x_{\br'},\qquad&\text{if }\br'=\br\pm\hate_x,
\\*[2mm]
\CalS^{-2}\,S^y_{\br}S^y_{\br'},\qquad&\text{if }\br'=\br\pm\hate_y.
\end{cases}
\end{equation}
\item[(5)]
The 120-degree model on~$\Z^3$:
\begin{equation}
\label{H5}
H=\sum_{\langle\br,\br'\rangle}\CalS^{-2}\,T_{\br}^jT_{\br'}^j\quad\text{if}\quad\br'=\br\pm\hate_j
\end{equation}
where
\begin{equation}
T_{\br}^j=
\begin{cases}
S_{\br}^x,\qquad&\text{if }j=1,
\\*[2mm]
-\tfrac12S_{\br}^x+\tfrac{\sqrt3}2S_{\br}^y,\qquad&\text{if }j=2,
\\*[2mm]
-\tfrac12S_{\br}^x-\tfrac{\sqrt3}2S_{\br}^y,\qquad&\text{if }j=3.
\end{cases}
\smallskip
\end{equation}
\end{enumerate}
Here~$\langle\br,\br'\rangle$ denotes a nearest-neighbor pair on~$\Z^d$---where unless specified we are only assuming~$d\ge2$---the symbol $\hate_j$ stands for the unit vector in the~$j$-th lattice direction and $\bS_{\br}=(S^x_{\br},S^y_{\br},S^z_{\br})$ is a triplet of spin-$\CalS$ operators for the spin at site~$\br$. The scaling of all interactions by the indicated
inverse powers of~$\CalS$ 
is necessary to make the~$\CalS\to\infty$ limit meaningful.

Model~(1) has been included only for illustration; the requisite transition was proved for large anisotropy~\cite{FL} and, in the context of the ferromagnet (which is not even reflection positive), for arbitrarily small anisotropy~\cite{Kennedy}. The classical versions of models (2-4) feature strong order-disorder transitions at intermediate temperatures; cf~\cite{Dobrushin-Shlosman,Kotecky-Shlosman,Alexander-Chayes,vES}. Here we will prove that corresponding transitions occur for large-$\CalS$ quantum versions of these systems. Models~(4-5) are quite unusual even at the classical level: Notwithstanding the fact that the 
Hamiltonian has only discrete symmetries, there is a \emph{continuum} of ground states. As was shown in~\cite{BCN1,BCN2}, at positive temperatures the degeneracy is lifted leaving only a finite number of preferential directions. The proofs of~\cite{BCN1,BCN2} involve (classical) spin-wave calculations not dissimilar to those of~\cite{Dyson1,Dyson2}. However, since the massless spin-wave excitations are central to the behavior of these systems---even at the classical level---it is by no means clear how to adapt the methods of~\cite{DLS,FL,FILS1,FILS2,Kennedy,Kotecky-Shlosman,BKU,DFF1,DFF2,KU} to these cases. 

The remainder of the paper is organized as follows: In the next section, we recall the formalism of coherent states, which is the basis of many $\CalS\to\infty$ limit results, and the techniques of reflection positivity and chessboard estimates, which underline many proofs of phase transitions in quantum systems. In Sect.~\ref{sec3} we state our main theorems; the proofs come in Sect.~\ref{sec4}. Applications to the various phase transitions in the aforementioned models are the subject of Sect.~\ref{sec5}. The Appendix (Sect.~\ref{sec6}) contains the proofs of some technical results that would 
detract from the main line of argument in Sects.~\ref{sec5.3}-\ref{sec5.5}.

\section{Preliminaries}
\label{sec2}\noindent
In this section, we summarize standard and well-known facts about
the $\textrm{SU}(2)$ coherent states (Sect.~\ref{sec2.1}) and the techniques of chessboard estimates (Sect.~\ref{sec2.2}). The purpose of this section is mostly informative; a reader familiar with these concepts may skip this section altogether and pass directly to the statement of main results in Sect.~\ref{sec3}.

\subsection{Coherent states}
\label{sec2.1}\noindent
Here we will recall the Bloch coherent states which were the basis for rigorous control of various classical limits of quantum spin systems~\cite{Berezin,Lieb,Fuller-Lenard1,Fuller-Lenard2,Simon}. In a well defined sense, these states are the ``closest'' objects to classical states that one can find in the Hilbert space. Our presentation follows closely Lieb's article~\cite{Lieb}; some of the calculations go back to~\cite{Arecchi}. The theory extends to general compact Lie groups, see~\cite{Simon,Duffield} for results at this level of generality. The literature on the subject of coherent states is quite large; we refer to, e.g.,~\cite{Perelomov,Ali-et-al} for comprehensive review and further references.

\smallskip
Given~$\CalS\in\{\ffrac12,1,\ffrac32,\dots\}$, consider the $(2\CalS+1)$-dimensional irreducible representation of the Lie algebra $\mathfrak{su}(2)$.
The generators, $(S^x,S^y,S^z)$, obeying the commutation rules~$[S^i,S^j]=2\texti\varepsilon_{ijk}S^k$, are operators acting on $\text{span}\{\ket M\colon M=-\CalS,\CalS+1,\dots,\CalS-1,\CalS\}\simeq\C^{2\CalS+1}$. In terms of spin-rasing/lowering operators,~$S^\pm=S^x\pm\texti S^y$, we have
\begin{equation}
\label{S-repre}
\begin{aligned}
S^z\ket M&=M\,\ket M,\\
S^+\ket M&=\sqrt{\CalS(\CalS+1)-M(M+1)}\,\ket{M+1},\\
S^-\ket M&=\sqrt{\CalS(\CalS+1)-M(M-1)}\,\ket{M-1}.
\end{aligned}
\end{equation}
In particular, $S^x$ and~$S^z$ are real while~$S^y$ is purely imaginary.

The classical counterpart of $\mathfrak{su}(2)$-spins are vectors on the two-dimensional unit sphere $\scrS_2$ in~$\R^3$. For each $\Omega\in\scrS_2$, one defines the coherent state vector in the direction $\Omega$ to be
\begin{equation}
\label{Omega-def}
    \ket{\Omega}= \sum_{M=-\CalS}^{\CalS}
    \binom{2\CalS}{\CalS+M}^{\ffrac12}\,\,
    [\cos(\ffrac\theta2)]^{\CalS+M}\,
    [\sin(\ffrac\theta2)]^{\CalS-M}\,
    \texte^{\texti(\CalS-M)\phi}\, \ket{M}.
\end{equation}
Here $(\theta,\phi)$ are the spherical coordinates of $\Omega$, with $\theta$ denoting the azimuthal angle and $\phi$ denoting the polar angle.
Let~$\zeta = \tan(\ffrac\theta2)\texte^{\texti\phi}$ denote the stereographic projection from $\scrS_2$ to $\C$. Then \eqref{Omega-def} can be written as 
\begin{multline}
\label{ket-Omega}
\qquad
\ket{\Omega}\, =\, \texte^{\,\zeta S^--\bar\zeta S^+}\ket\CalS=
[1+|\zeta|^2]^{-\CalS}\,\texte^{\zeta S^-}\, \ket{\CalS}
\\
=[\cos(\ffrac\theta2)]^{2\CalS}\, \exp(\tan(\ffrac\theta2)\texte^{\texti \phi} S^-)\, \ket{\CalS}\, .
\qquad
\end{multline}
One important property of the coherent state $\ket\Omega$ is that it is an eigenvector of the matrix
$\Omega \cdot \bS$ with maximal eigenvalue:
\begin{equation}
\label{eq:coherenteig}
(\Omega \cdot \bS) \ket{\Omega}\, =\, \CalS \ket{\Omega}\, .
\end{equation}
This equation characterizes the vector $\ket{\Omega}$ up to a phase factor.
The choice of the phase factors may seem arbitrary, but
in practice they will cancel in all the formulas we use.

The fact that the states~$\ket\Omega$ have been defined relative to the basis in \eqref{S-repre} is inconsequential. Indeed, a rotation of a coherent state is, to within a harmless phase factor, the coherent state corresponding to the rotated vector. 
More precisely, for each $\omega \in \scrS_2$ and $t  \in \R$, one may consider the unitary $U_{\omega,t} =\texte^{\texti t (\omega \cdot \bS)}$.
Then, for any $\Omega \in \scrS_2$, a simple calculation shows that
\begin{equation}
U_{\omega,t} (\Omega \cdot \bS) U_{\omega,t}^+\, =\, R_{\omega,t}(\Omega)\cdot \bS\, ,
\end{equation}
where $R_{\omega,t} \in \text{\rm SO}(3)$ is the rotation about the ray passing through $\omega$ by the angle $t$.
Because of this $U_{\omega,t} \ket{\Omega}$ satisfies (\ref{eq:coherenteig}) with $\Omega$ replaced
by $R_{\omega,t}(\Omega)$ and so 
$$
U_{\omega,t} \ket{\Omega}\, =\, \texte^{\texti f(\Omega,\omega,t)}\, \ket{R_{\omega,t}(\Omega)}\, ,
$$
for some phase factor $f(\Omega,\omega,t)$.
Since SU(2) is a double cover of SO(3), $f(\Omega,\omega,2\pi)$ is not necessarily 0 (mod $2\pi$); rather $\texte^{\texti f(\Omega,\omega,2\pi)} = (-1)^{2\CalS}$.

\smallskip
The explicit formula \eqref{Omega-def} for~$\ket\Omega$ yields
\begin{equation}
\ip{\Omega'}{\Omega}= \bigl[\cos(\ffrac\theta2) \cos(\ffrac{\theta'}2) + \texte^{\texti(\phi-\phi')}\sin(\ffrac\theta2) \sin(\ffrac{\theta'}2)\bigr]^{2\CalS}.
\end{equation}
Defining the angle between $\Omega$ and $\Omega'$ to be $\Theta$,
one also has
\begin{equation}
\label{overlap}
    \bigl|\ip{\Omega'}{\Omega}\bigr|= [\cos(\ffrac\Theta2)]^{2\CalS}.
\end{equation}
Another formula that is directly checked from~\eqref{Omega-def} is
\begin{equation}
\label{resolution}
    \1= \frac{2\CalS+1}{4\pi} \int_{\scrS_2}
    \textd\Omega\, \proj\Omega,
\end{equation}
where~$\textd\Omega$ denotes the uniform surface measure on~$\scrS_2$ with total mass~$4\pi$.

Given any operator $A$ on $\C^{2\CalS+1}$, one can form what
is commonly known as the \emph{lower symbol}, which is a function $\Omega
\mapsto \langle A\rangle_\Omega$ defined by
\begin{equation}
\label{lower-symbol1}
    \langle A\rangle_\Omega\,:=\, \bra{\Omega} A \ket{\Omega}.
\end{equation}
(Here and henceforth, $\bra{\Omega} A \ket{\Omega}$ denotes the inner-product of $\ket{\Omega}$ with the
vector $A\ket{\Omega}$.) 
While not entirely obvious, it turns out that the trace of~$A$ admits the formula
\begin{equation}
\label{trace1}
\TR(A)=\frac{2\CalS+1}{4\pi} \int_{\scrS_2}
    \textd\Omega\, \langle A\rangle_\Omega.
\end{equation}
There is also a generalization of~\eqref{resolution}:
There exists a function $\Omega \mapsto [A]_\Omega$ such
that
\begin{equation}
\label{upper-symbol1}
    A= \frac{2\CalS+1}{4\pi} \int_{\scrS_2}
    \textd\Omega\, [A]_\Omega\, \proj\Omega.
\end{equation}
Any such $\Omega \mapsto [A]_\Omega$ is called an \emph{upper symbol} for~$A$. 
Unfortunately, such a function is not unique and so $[A]_\Omega$
actually represents an equivalence class of functions. 
Obviously $\langle A+B\rangle_\Omega = \langle A\rangle_\Omega
+ \langle B\rangle_\Omega$. For the upper symbols, if
$[A]_\Omega$ and $[B]_\Omega$ are upper symbols for $A$ and
$B$ then $[A+B]_\Omega = [A]_\Omega + [B]_\Omega$ is an
upper symbol for $A+B$.

When $A=\1$, one has $\langle\1\rangle_\Omega=1$ and,
by~\eqref{resolution}, one can also choose
$[A]_\Omega=1$. However, it is usually not the case that the lower symbol
is also an upper symbol, e.g., we have
\begin{equation}
\label{upper-lower}
\begin{aligned}
\langle S^x \rangle_\Omega&= \CalS\, \sin\theta\, \cos\phi,\\  
\langle S^y \rangle_\Omega&= \CalS\, \sin\theta\, \sin\phi,\\
\langle S^z \rangle_\Omega&= \CalS\, \cos\theta,\\
\end{aligned}
\qquad\qquad
\begin{aligned}
\ 
[S^x]_\Omega&= (\CalS+1)\, \sin\theta\, \cos\phi,\\
[S^y]_\Omega&= (\CalS+1)\, \sin\theta\, \sin\phi,\\
[S^z]_\Omega&= (\CalS+1)\, \cos\theta.\\
\end{aligned}
\end{equation}
As is easily checked, the leading order in~$\CalS$ of these expressions is exactly the classical counterpart of the corresponding operator. 
For more complicated products of the spin components, both symbols develop lower-order ``non-classical'' corrections but, as was shown in~\cite[Theorem~2]{Duffield}, the leading order term is always the classical limit.

\smallskip
The above formalism generalizes to collections of many spins. Let~$\Lambda$ be a finite set
and, for each~$\br\in\Lambda$, let~$(S_{\br}^1,S_{\br}^2,S_{\br}^3)$ be the spin operator for the spin at~$\br$. 
We will assume that the spins at all sites have magnitude $\CalS$, so we assume to have a joint (product)
representation of these spins on~$\HH_{\Lambda} = \bigotimes_{\br\in
\Lambda} [\C^{2\CalS+1}]_r$. Consider an assignment of
a classical spin~$\Omega_{\br}\in\scrS_2$ to each $\br\in \Lambda$ and
denote the resulting configuration $(\Omega_{\br})_{\br\in\Lambda}$ by~$\Omega$.
The desired product coherent state then is
\begin{equation}
    \ket{\Omega}\,:=\, \bigotimes_{\br\in \Lambda}
    \ket{\Omega_{\br}}.
\end{equation}
Given an operator~$A$ on $\HH_{\Lambda}$, we define its lower symbol by the generalization~of~\eqref{lower-symbol1},
\begin{equation}
\label{lower-symbol}
    \langle A\rangle_\Omega= \bra{\Omega}A\ket{\Omega},
    \qquad\Omega\in(\scrS_2)^{|\Lambda|}.
\end{equation}
With this lower symbol we may generalize \eqref{trace1} into
\begin{equation}
\label{trace}
\TR_{\HH_\Lambda}(A)=\left(\frac{2\CalS+1}{4\pi}\right)^{|\Lambda|}
    \int_{(\scrS_2)^{|\Lambda|}} \textd\Omega\,\,\langle A\rangle_\Omega.
\end{equation}
There is also a representation of~$A$ in terms of an upper symbol~$[A]_\Omega$,
\begin{equation}
\label{upper-symbol}    
	A=\left(\frac{2\CalS+1}{4\pi}\right)^{|\Lambda|}
    \int_{(\scrS_2)^{|\Lambda|}} \textd\Omega\,\,[A]_\Omega\,
    \proj\Omega,
\end{equation}
where~$\textd\Omega$ is the product surface measure on~$(\scrS_2)^{|\Lambda|}$ and where~$\Omega\mapsto[A]_\Omega$ is now a function $(\scrS_2)^{|\Lambda|}\to\C$. A special case of this formula is the resulution of the identity on~$\HH_\Lambda$. Note that \eqref{upper-symbol} allows us to substitute $[A]_\Omega$ for $\langle A\rangle_\Omega$ in~\eqref{trace}.

It is easy to check that~$\Omega\mapsto[A]_\Omega$ has the expected behavior under (tensor) product of operators, provided these respect the product structure of~$\HH_\Lambda$. Indeed, suppose that~$\Lambda$ is the disjoint union of~$\Lambda_1$ and~$\Lambda_2$ and let~$\ket{\Omega_1}$ and~$\ket{\Omega_2}$ be product coherent states from~$\HH_{\Lambda_1}$ and~$\HH_{\Lambda_2}$, respectively. Given two operators~$A_1\colon\HH_{\Lambda_1}\to\HH_{\Lambda_1}$ and $A_2\colon\HH_{\Lambda_2}\to\HH_{\Lambda_2}$, let~$[A_1]_{\Omega_1}$ and~$[A_2]_{\Omega_2}$ be their associated upper symbols. Then
\begin{equation}
\label{2.11}
[A_1\otimes A_2]_{(\Omega_1,\Omega_2)}\,:=\,[A_1]_{\Omega_1}\,[A_2]_{\Omega_2}
\end{equation}
is an upper symbol of $A_1\otimes A_2$ relative to state $\ket{(\Omega_1,\Omega_2)}=\ket{\Omega_1}\otimes\ket{\Omega_2}$.
On the other hand, if~$[A]_\Omega$ depends only on~$(\Omega_{\br})_{\br\in\Lambda'}$ where~$\Lambda'\subsetneqq\Lambda$, then we can perform a partial trace in~\eqref{upper-symbol} by integrating over the~$(\Omega_{\br})_{\br\in\Lambda\smallsetminus\Lambda'}$ and applying~\eqref{resolution} for each integral.

\subsection{Chessboard estimates}
\label{sec2.2}\noindent
Next we will review the salient features of the technology of reflection positivity/chessboard estimates which was developed and applied to both classical and quantum systems in the works of F.~Dyson, J.~Fr\"ohlich, R.~Israel, E.~Lieb, B.~Simon and T.~Spencer~\cite{FSS,DLS,FL,FILS1,FILS2}.

\smallskip
Consider a $C^\star$-algebra~$\frakA$ and suppose that~$\frakA_+$ and~$\frakA_-$ are \emph{commuting} subalgebras which are ``mirror images'' of each other in the sense that there is an algebraic automorphism~$\theta\colon\frakA\to\frakA$ such that~$\theta(\frakA_\pm)=\frakA_\mp$ and~$\theta^2=\text{id}$. Assuming that~$\frakA$ is represented in terms of complex matrices, for~$A\in\frakA$ we define~$\bar A$ to be the complex conjugate---\emph{not} the adjoint---of~$A$. We will always assume that~$\frakA$ is closed under complex conjugation. Note that, since complex conjugation is not a ``covariant operation,'' the representation of~$\frakA$ ought to stay fixed throughout all calculations involving complex conjugation.

A relevant example of the above setting is a quantum spin-$\CalS$ system on the $d$-dimensional torus~$\T_L$ of~$L\times\dots\times L$ sites, with~$L$ even, which we think of as a union of two \emph{disjoint} symmetric halves, $\T_L^+$ and~$\T_L^-$. 
(Note that $\T_L$ can also be identified with $\Z^d/L\Z^d$.
Of course the origin $0 \in \Z^d$ maps to the origin of the torus.)
Then~$\frakA$ is the $C^\star$-algebra of all observables---represented by $(2\CalS+1)^{|\T_L|}$ dimensional complex matrices---and~$\frakA_\pm$ are the sets of observables on~$\T_L^\pm$, respectively. Explicitly,~$\frakA_+$ are matrices of the form~$A_+\otimes\1$, where~$A_+$ ``acts'' only on~$\T_L^+$, while the matrices in~$\frakA_-$ take the form~$\1\otimes A_-$. The operation~$\theta$ is the map that interchanges the ``left'' and ``right'' half of the torus; e.g., in a properly parametrized basis, $\theta(A_+\otimes\1)=\1\otimes A_+$. The fact that~$\theta$ arises from a reflection leads to the following concept:

\begin{mydefinition}
Let~$\langle-\rangle$ be a state---i.e., a continuous linear functional---on~$\frakA$ and let~$\theta$ be as above. We say that~$\langle-\rangle$ is \emph{reflection positive} (relative to~$\theta$) if for all~$A,B\in\frakA_+$,
\begin{equation}
\bigl\langle A\,\overline{\theta(B)}\bigr\rangle
=\overline{\bigl\langle B\,\overline{\theta(A)}\bigr\rangle}
\end{equation}
and
\begin{equation}
\label{RP}
\bigl\langle A\,\overline{\theta(A)}\bigr\rangle\ge0.
\end{equation}
\end{mydefinition}

The following condition, derived in~\cite[Theorem~E.1]{DLS} and in~\cite[Theorem~2.1]{FL}, is sufficient for the Gibbs state to have the above property: 

\begin{theorem}[Reflection positivity---sufficient condition]
\label{thm-RP}
Given a reflection of~$\T_L$ as described above and using~$\theta$ to denote the associated reflection operator, if the Hamiltonian of a quantum system on~$\T_L$ can be written as
\begin{equation}
\label{RP-Ham}
H=C+\overline{\theta(C)}-\int\varrho(\textd\alpha)\, D_\alpha\,\overline{\theta(D_\alpha)},
\end{equation}
where~$C,D_\alpha\in\frakA_+$ and~$\varrho$ is a (finite) positive measure, then the canonical Gibbs state $\state{-}_{L,\beta}$, which is defined by
\begin{equation}
\label{thermal-state}
\state{A}_{L,\beta}=\frac{\TR_{\HH_{\T_L}}(\texte^{-\beta H}A)}{\TR_{\HH_{\T_L}}(\texte^{-\beta H})},
\end{equation}
is reflection positive relative to~$\theta$ for all~$\beta\ge0$.
\end{theorem}

The crux of the proof of \eqref{RP} is the fact that the $\beta=0$ state is \emph{generalized reflection positive}, i.e., $\langle A_1\overline{\theta(A_1)}\dots A_n\overline{\theta(A_n)}\rangle_{L,0}\ge0$. The rest follows by a Lie-Trotter expansion of~$\texte^{-\beta H}$ into powers of the last term in \eqref{RP-Ham}---hence the need for a \emph{minus} sign in front of the integral.

\begin{myremark}
\label{no-quantum-CE}
We reiterate that the reflections of~$\T_L$ considered here are always for ``planes of reflections'' \emph{between} sites. In classical models one can also consider the (slightly more robust) reflections for ``planes'' on sites. However, due to non-commu\-ta\-tivity issues, Theorem~\ref{thm-RP} does not seem to generalize to quantum systems for these kinds of reflections.
\end{myremark}

Reflection positivity has two important (and related) consequences: \emph{Gaussian domination}---leading ultimately to infrared bounds---and \emph{chessboard estimates}. In this work we make no use of the former; we proceed by discussing the details of the latter.

\smallskip
Let~$\Lambda_B$ be a block of~$B\times\dots\times B$ sites with the ``lower-left'' corner at the origin. Assuming that~$L$ is a multiple of~$B$, we can tile~$\T_L$ by disjoint translates of~$\Lambda_B$. The positions of these translates are given by $B$-multiples of vectors~$\bt$ from the factor torus~$\T_{L/B}$. In particular, if~$\Lambda_B+\br$ denotes the translate of~$\Lambda_B$ by~$\br\in\T_L$, then~$\T_L$ is the disjoint union~$\bigcup_{\bt\in\T_{L/B}}(\Lambda_B+B\bt)$.
Let~$\frakA_{\Lambda_B}$ denote the algebra of observables in~$\Lambda_B$, i.e., each~$A\in\frakA_{\Lambda_B}$ has the form~$A=A_B\otimes\1$, where~$A_B$ acts only on the portion of the Hilbert space corresponding to~$\Lambda_B$. For each~$A\in\frakA_{\Lambda_B}$ and each $\bt  \in \T_{L/B}$ with~$|\bt|=1$, we can define an antilinear operator~$\whvartheta_{\bt}(A)$ in~$\Lambda_B+B\bt$~by
\begin{equation}
\whvartheta_{\bt}(A)=\overline{\theta(A)}
\end{equation}
where~$\theta$ is the operator of reflection along the corresponding side of~$\Lambda_B$. By taking further reflections, we can define~$\whvartheta_{\bt}(A)$ for every~$\bt\in\T_{L/B}$. 
(Thus~$\whvartheta_{\bt}$ is linear for even-parity~$\bt$ and antilinear for odd-parity~$\bt$; if every component of~$\bt$ is even then $\whvartheta_{\bt}$ is simply the
translation by~$B\bt$.)
It is easy to check that the resulting~$\whvartheta_{\bt}(A)$ does not depend on what sequence of reflections has been used to generate it.

\smallskip
The fundamental consequence of reflection positivity, derived in a rather general form in~\cite[Theorem~2.2]{FL}, is as follows:

\begin{theorem}[Chessboard estimate]
\label{thm-CE}
Suppose that the state~$\state{-}$ is reflection positive for any ``plane of reflection'' between sites on~$\T_L$. Then for any~$A_1,\dots,A_m\in\frakA_{\Lambda_B}$ and any \emph{distinct} vectors $\bt_1,\dots,\bt_m\in\T_{L/B}$,
\begin{equation}
\label{chessboard}
\state{\,\prod_{j=1}^m\whvartheta_{\bt_j}(A_j)}
\le\,\prod_{j=1}^m\,\state{\,\,\prod_{\bt\in\T_{L/B}}\whvartheta_{\bt}(A_j)}^{(B/L)^d}.
\end{equation}
\end{theorem}

By \eqref{chessboard} we may bound the expectation of a product of operators by product of expectations of so called ``disseminated'' operators. As we will show on explicit examples later, these are often easier to estimate. Note that the giant products above can be written in any order by our assumption that the block-operators in different blocks commute. 

A corresponding statement works also for classical reflection-positive measures. The only formal difference is  that  the~$A_j$'s are replaced by functions, or indicators of events~$\AA_j$, which  depend only on the spin configuration in~$\Lambda_B$.  Then equation \eqref{chessboard} becomes
\begin{equation}
\label{class-chessboard}
\BbbP\biggl(\,\bigcap_{j=1}^m\theta_{\bt_j}(\AA_j)\biggr)
\le\,\prod_{j=1}^m\,\BbbP\biggl(\,\bigcap_{\bt\in\T_{L/B}}\theta_{\bt}(\AA_j)\biggr)^{(B/L)^d}.
\end{equation}
Here~$\theta_{\bt}(\AA)$ is the (usual) reflection of~$\AA$ to the block~$\Lambda_B+B\bt$. (We reserve the symbol~$\vartheta_{\bt}(\AA)$ for an operation that more closely mimics~$\whvartheta_{\bt}$ in the coherent-state representation; see the definitions right before Proposition~\ref{prop3.4}.)
Refs.~\cite{BCN1,BCKiv,BK} contain a detailed account of the above formalism in the classical context; the original statements are, of course, due to~\cite{FL,FILS1,FILS2}.

\begin{myremark}
\label{RP-peculiar}
Unlike its classical counterpart, the quantum version of reflection positivity is a rather mysterious concept. First, for most of the models listed in the introduction, in order to bring the Hamiltonian to the form \eqref{RP-Ham}, we actually have to perform some sort of rotation of the spins. (We may think of this as choosing a different representation of the spin operators.) The purpose of this operation is to have all spins ``represented'' by real-valued matrices, while making the overall sign of the interactions negative. This permits an application of Theorem~\ref{thm-RP}. 

It is somewhat ironic that this works beautifully for antiferromagnets, which thus become effectively ferromagnetic, but fails miserably~\cite{Speer} for genuine ferromagnets. For~XY-type models, when only two of the spin-components are involved in the interaction, we can always choose a representation in which all matrices are real valued. If only quadratic interactions are considered (as for the nematics) the overall sign is inconsequential but, once interactions of different degrees are mixed---even if we just add a general external field to the Hamiltonian---reflection positivity may fail again.
\end{myremark}

\section{Main results}
\label{sec3}\noindent
We now give precise statements of our main theorems. First we will state a bound on the matrix elements of the Gibbs-Boltzmann weight in the (overcomplete) basis of coherent states. On the theoretical side, this result generalizes the classic Berezin-Lieb inequalities~\cite{Berezin,Lieb} and thus provides a more detailed demonstration of the approach to the classical limit as~$\CalS\to\infty$. On the practical side, the bound we obtain allows us to replace the ``exponential localization'' technique of Fr\"ohlich and Lieb~\cite{FL}---which is intrinsically quantum---by an estimate for the classical version of the model. 

The rest of our results show in detail how Theorem~\ref{thm-matrix-element} fits into the standard line of proof of phase transitions via chessboard estimates. In Sect.~\ref{sec5} we will apply this general strategy to the five models of interest.

\subsection{Matrix elements of Gibbs-Boltzmann weights}
We commence with a definition of the class of models to which our arguments apply.
Consider a finite set~$\Lambda\subset\Z^d$ and, for each~$\Gamma\subset\Lambda$, let~$h_\Gamma$ be an operator on~$\HH_\Lambda=\bigotimes_{\br\in\Lambda}[\C^{2\CalS+1}]_{\br}$ that depends only on the spins in~$\Gamma$. (I.e.,~$h_\Gamma$ is a tensor product of an operator on~$\HH_\Gamma$ and the unity on~$\HH_{\Lambda\smallsetminus\Gamma}$.) We will assume that~$h_\Gamma=0$ if the size of $\Gamma$ exceeds some finite constant, i.e., each interaction term involves only a bounded number of spins. The Hamiltonian is then
\begin{equation}
\label{Hamiltonian}
H=\sum_{\Gamma\colon\Gamma\subset\Lambda}h_\Gamma.
\end{equation}
Most of the interesting examples are such that~$h_\Gamma=0$ unless~$\Gamma$ is a two point set $\{x,y\}$ containing a pair of nearest neighbors on~$\Z^d$---as is the case of all of the models~(1-5) discussed in Sect.~\ref{sec1}.

As already noted, our principal technical result is a bound on the matrix element $\langle\Omega|\texte^{-\beta H}|\Omega'\rangle$. To state this bound precisely, we need some more notation. Let~$\Omega\mapsto[h_\Gamma]_\Omega$ be an upper symbol of the operator~$h_\Gamma$ which, by~\eqref{2.11}, may be assumed independent of the components $(\Omega_{\br})_{x\not\in\Gamma}$. We \emph{fix} the upper symbol of~$H$ to
\begin{equation}
\label{H-upper}
[H]_\Omega=\sum_{\Gamma\colon\Gamma\subset\Lambda}[h_\Gamma]_\Omega.
\end{equation}
We will also use~$|\Gamma|$ to denote the number of elements in the set~$\Gamma$ and $\Vert h_\Gamma\Vert$ to denote the operator norm of~$h_\Gamma$ on~$\HH_\Lambda$.

Let~$|\Omega_{\br}-\Omega_{\br}'|$ denote the (3-dimensional) Euclidean distance of the points~$\Omega_{\br}$ and~$\Omega_{\br}'$ on~$\scrS_2$, and consider the following $\ell^1$ and~$\ell^2$-norms on~$(\scrS_2)^{|\Lambda|}$:
\begin{equation}
\norm\Omega{\Omega'}_1=\sum_{\br\in\Lambda}|\Omega_{\br}-\Omega_{\br}'|
\end{equation}
and
\begin{equation}
\norm\Omega{\Omega'}_2=\biggl(\,\sum_{\br\in\Lambda}|\Omega_{\br}-\Omega_{\br}'|^2\biggr)^{\ffrac12}.
\end{equation}
Besides these two norms, we will also need the ``mixed'' quantity
\begin{equation}
\dall{\Omega}{\Omega'}=\sum_{\br\in\Lambda}\bigl(\sqrt{\CalS}|\Omega_{\br}-\Omega_{\br}'|\wedge\CalS|\Omega_{\br}-\Omega_{\br}'|^2\bigr),
\end{equation}
where~$\wedge$ denotes the minimum.
This is not a distance function but, as will be explained in Lemma~\ref{lemma-triangle}, it does satisfy an inequality
which could be compared to the triangle inequality.
Finally, from~\eqref{overlap} we know that~$|\ip{\Omega_{\br}}{\Omega_{\br}'}|=1-O(\CalS|\Omega_{\br}-\Omega_{\br}'|^2)$. Hence, there is~$\eta>0$ such that
\begin{equation}
\label{d-eta}
\bigl|\langle\Omega|\Omega'\rangle\bigr|\le\texte^{-\eta\dtwo\Omega{\Omega'}}
\end{equation}
holds for all~$\CalS$, all~$\Omega,\Omega'\in(\scrS_2)^{|\Lambda|}$ and all~$\Lambda$. We fix this~$\eta$ throughout all forthcoming derivations.
(Since $[\cos(\ffrac\Theta2)]^2 = 1 - \ffrac{1}{4}\|\Omega-\Omega'\|^2$ for a single spin, we have $\eta=\ffrac14$. But $\eta$ plays only a marginal role in our calculations so we will leave it implicit.)
 Our first main theorem then is:

\begin{theorem}
\label{thm-matrix-element}
Suppose that there exists a number~$R$ such that
\begin{equation}
\label{range}
|\Gamma|>R\quad\Rightarrow\quad h_\Gamma=0,
\end{equation}
and that, for some constants~$c_0$ and $c_1$ independent of~$\CalS$ and~$\Lambda$, we have
\begin{equation}
\label{uniform-bd}
\sup_{x \in \Lambda}\, \sum_{\Gamma\colon\\ x\in\Gamma\subset\Lambda}\Vert h_\Gamma\Vert\le c_0
\end{equation}
as well as the Lipschitz bound
\begin{equation}
\label{Lipschitz}
\bigl|[h_\Gamma]_\Omega-[h_\Gamma]_{\Omega'}\bigr|\le c_1\norm\Omega{\Omega'}_1\Vert h_\Gamma\Vert,
\qquad \Gamma\subset\Lambda.
\end{equation}
Then for any constant~$c_2>0$, there exists a constant~$c_3>0$, depending only on~$c_0$, $c_1$, $c_2$ and~$R$, such that for all~$\beta\le c_2\sqrt{\CalS}$,
\begin{equation}
\label{element}
\bigl|\langle\Omega|\texte^{-\beta H}|\Omega'\rangle\bigr|
\le\texte^{-\beta[H]_\Omega-\eta\dall{\Omega}{\Omega'}+c_3\beta|\Lambda|/\sqrt{\CalS}}
\end{equation}
holds for all~$\Omega,\Omega'\in(\scrS_2)^{|\Lambda|}$ and all finite~$\Lambda$.
\end{theorem}

Note that we do not assume that the Hamiltonian is translation-invariant. In fact, as long as the conditions \twoeqref{range}{Lipschitz} hold as stated, the geometry of the underlying set is completely immaterial. For the diagonal elements---which is all we need in the subsequent derivations anyway---the above bound becomes somewhat more transparent:

\begin{mycorollary}
\label{cor3.2}
Suppose \twoeqref{range}{Lipschitz} hold and let~$c_2$ and~$c_3$ be as in Theorem~\ref{thm-matrix-element}. Then for all $\beta$ and~$\CalS$ with~$\beta\le c_2\sqrt S$, all~$\Omega\in(\scrS_2)^{|\Lambda|}$ and all~$\Lambda$,
\begin{equation}
\label{3.11}
\texte^{-\beta\langle H\rangle_\Omega}\le
\langle\Omega|\texte^{-\beta H}|\Omega\rangle\le
\texte^{-\beta[H]_\Omega+c_3\beta|\Lambda|/\sqrt{\CalS}}
\end{equation}
\end{mycorollary}

It is interesting to compare this result with the celebrated Berezin-Lieb inequalities~\cite{Berezin,Lieb} which state the following bounds between quantum and classical partition functions:
\begin{equation}
\label{3.12a}
\int_{(\scrS_2)^{|\Lambda|}}\frac{\textd\Omega}{(4\pi)^{|\Lambda|}}\,\,
\texte^{-\beta\langle H\rangle_\Omega}
\le
\frac{\TR_{\HH_\Lambda}(\texte^{-\beta H})}{(2\CalS+1)^{|\Lambda|}}
\le
\int_{(\scrS_2)^{|\Lambda|}}\frac{\textd\Omega}{(4\pi)^{|\Lambda|}}\,\,
\texte^{-\beta[H]_\Omega}.
\end{equation}
(An unpublished proof of E.~Lieb, cf~\cite{Simon-book}, shows both inequalities are simple consequences of Jensen's inequality; the original proof~\cite{Lieb} invoked also the ``intrinsically non-commutative'' Golden-Thompson inequality.)
From Corollary~\ref{cor3.2} we now know that, to within a correction of order~$\beta/\sqrt S$, the estimates corresponding to \eqref{3.12a} hold even for the (diagonal) matrix elements relative to coherent states. However, the known proofs of \eqref{3.12a} use the underlying trace structure in a very essential way and are not readily extended to a generalization along the lines of \eqref{3.11}.

\begin{myremarks}
Some comments are in order:
\settowidth{\leftmargini}{(11)}
\begin{enumerate}
\item[(1)]
The correction of order~$\beta|\Lambda|/\sqrt\CalS$ is the best one can do at the above level of generality. Indeed, when~$\Omega$ and~$\Omega'$ are close in the sense~$\norm\Omega{\Omega'}_1=O(|\Lambda|/\sqrt\CalS)$, then $[H]_\Omega$ and~$[H]_{\Omega'}$ differ by a quantity of order~$c_1|\Lambda|/\sqrt\CalS$. Since the matrix element is symmetric in~$\Omega$ and~$\Omega'$, the bound must account for the difference. However, there is a deeper reason why~$\beta/\sqrt\CalS$ needs to be small for the classical Boltzmann weight to faithfully describe the matrix elements of the quantum Boltzmann weight. Consider a single spin with the Hamiltonian $H=\CalS^{-1}S^z$, and let~$\Omega$ correspond to the spherical angles~$(\theta,\phi)$. A simple calculation shows that then
\begin{equation}
\begin{aligned}
\langle\Omega|\texte^{-\beta H}|\Omega\rangle
&=\bigl[\cos^2(\ffrac\theta2)\texte^{-\frac12\beta/\CalS} 
+\sin^2(\ffrac\theta2)\texte^{\frac12\beta/\CalS}\bigr]^{2\CalS}
\\
&=\texte^{-\beta\cos\theta+\frac{\beta^2}{4\CalS}(1-\cos^2\theta)+O(\beta^3/\CalS^2)}
\end{aligned}
\end{equation}
The term~$\beta\cos\theta$ is the (now unambiguous) classical interaction in ``state''~$\Omega$. The leading correction is of order~$\beta^2/\CalS$, which is only small if~$\beta\ll\sqrt\CalS$.
\item[(2)]
Another remark that should be made, lest the reader think about optimizing over the many choices of upper symbols in (\ref{element}): The constant~$c_3$ depends on the upper symbol.
For~$h_\Gamma$ being a polynomial in spin operators, $[h_\Gamma]$ may be chosen a polynomial too~\cite[Proposition~3]{Duffield}. This automatically ensures properties such as the Lipschitz continuity (as well as existence of the classical limit, cf~\eqref{3.14a}). For more complex $h_\Gamma$'s---e.g., those defined by an infinite power series---one must carefully check the conditions \twoeqref{range}{Lipschitz} before Theorem~\ref{thm-matrix-element} can be applied.
\end{enumerate}
\end{myremarks}

\subsection{Absence of clustering}
\noindent
Our next task is to show how Theorem~\ref{thm-matrix-element} can be applied to establish phase transitions in models whose ($\CalS\to\infty$) classical version exhibits a phase transition that can be proved by means of chessboard estimates. The principal conclusion is the \emph{absence of clustering} which, as we will see in Sect.~\ref{sec3.3}, directly implies a quantum phase transition.

\smallskip
Consider the setting as described in Sect.~\ref{sec2.2}, i.e., we have a torus~$\T_L$ of side~$L$ which is tiled by~$(L/B)^d$ disjoint translates of a block~$\Lambda_B$ of side~$B$. For each operator in~$\Lambda_B$ and each~$\bt\in\T_{L/B}$, we write~$\whvartheta_{\bt}(A)$ for the appropriate reflection---accompanied by complex conjugation if~$\bt$ is an odd parity site---of~$A$ ``into'' the block~$\Lambda_B+B\bt$. In addition to the operators on~$\HH_{\T_L}=\bigotimes_{\bt\in\T_L}[\C^{2\CalS+1}]_{\bt}$, we will also consider events~$\AA$ on the space of classical configurations~$(\scrS_2)^{|\T_L|}$ equipped with the Borel product~$\sigma$-algebra and the product surface measure~$\textd\Omega=\prod_{\br\in\T_L}\textd\Omega_{\br}$. If~$\AA$ is an event that depends only on the configuration in~$\Lambda_B$, we will call~$\AA$ a~\emph{$B$-block event}. For each~$\bt\in\T_{L/B}$, we use~$\theta_{\bt}(\AA)$ to denote the event in~$\Lambda_B+B\bt$ that is obtained by (pure) reflection of~$\AA$ ``into''~$\Lambda_B+B\bt$.

Given a quantum Hamiltonian~$H$ of the form \eqref{Hamiltonian}, let~$\langle-\rangle_{L,\beta}$ denote the thermal state~\eqref{thermal-state}. Considering the \emph{classical} Hamiltonian~$H^\infty\colon(\scrS_2)^{|\T_L|}\to\R$, which we define as 
\begin{equation}
\label{3.14a}
H^\infty(\Omega)=\lim_{\CalS\to\infty}\state{H}_\Omega=\lim_{\CalS\to\infty}[H]_\Omega,
\end{equation}
we use~$\BbbP_{L,\beta}$ to denote the usual Gibbs measure. 
Explictly, for any event~$\AA\subset(\scrS_2)^{|\T_L|}$,
\begin{equation}
\BbbP_{L,\beta}(\AA)=\int_\AA\textd\Omega\,\,\frac{\texte^{-\beta H^\infty(\Omega)}}{Z_L(\beta)},
\end{equation}
where~$Z_L(\beta)$ is the classical partition function. For each $B$-block event~$\AA$ we will also consider its disseminated version~$\bigcap_{\bt\in\T_{L/B}}\theta_{\bt}(\AA)$ and introduce the abbreviation
\begin{equation}
\label{frakp}
\frakp_{L,\beta}(\AA)=\biggl[\BbbP_{L,\beta}\Bigl(\,\bigcap_{\bt\in\T_{L/B}}\theta_{\bt}(\AA)\Bigr)\biggr]^{(B/L)^d}
\end{equation}
for the corresponding quantity on the right-hand side of \eqref{class-chessboard}.
An application of \eqref{chessboard} shows that~$\AA\mapsto\frakp_{L,\beta}(\AA)$ is an outer measure on the $\sigma$-algebra of~$B$-block events (cf~\cite[Theorem~6.3]{BCN1}).

For each measurable set~$\AA\subset(\scrS_2)^{|\T_L|}$ we consider the operator
\begin{equation}
\label{hatQ}
\hatQ_\AA=\left(\frac{2\CalS+1}{4\pi}\right)^{|\T_L|}
\int_\AA\textd\Omega\,\,\ket\Omega\bra\Omega.
\end{equation}
Since the coherent states are overcomplete, this operator is not a projection; notwithstanding, we may \emph{think} of it as  a non-commutative counterpart of the indicator of the event~$\AA$. In order to describe the behavior of~$\hatQ_\AA$ under~$\whvartheta_{\bt}$, we introduce the classical version~$\vartheta_{\bt}$ of~$\whvartheta_{\bt}$ which is defined as follows: Consider a ``complex-conjugation'' map~$\sigma\colon(\scrS_2)^{|\T_L|}\to(\scrS_2)^{|\T_L|}$ which, in a given representation of the coherent states, has the effect
\begin{equation}
\label{conjugate}
\overline{\proj\Omega}=\proj{\sigma\Omega}.
\end{equation}
For the representation introduced in Sect.~\ref{sec2.1}, we can choose~$\sigma$ to be the reflection through the~$xz$-plane (in spin space), i.e., if $\Omega=(\theta,\phi)$ then~$\sigma(\Omega)=(\theta,-\phi)$.
For even parity~$\bt\in\T_{L/B}$, we simply have~$\vartheta_{\bt}=\theta_{\bt}$ while for odd parity~$\bt\in\T_{L/B}$ we have~$\vartheta_{\bt}=\theta_{\bt}\circ\sigma$.

\smallskip
Here are some simple facts about the $\hatQ$-operators:

\begin{myproposition}
\label{prop3.4}
For any~$B$-block event~$\AA$ we have
\begin{equation}
\label{Q-theta}
\whvartheta_{\bt}(\hatQ_{\AA})=\hatQ_{\vartheta_{\bt}(\AA)},\qquad \bt\in\T_{L/B}.
\end{equation}
Moreover, if~$\AA_1,\dots,\AA_m$ are $B$-block events and~$\bt_1,\dots,\bt_m$ are \emph{distinct} elements from $\T_{L/B}$, then
\begin{equation}
\label{hatQ-commute}
[\hatQ_{\theta_{\bt_i}(\AA_i)},\hatQ_{\theta_{\bt_j}(\AA_j)}]=0,\qquad 1\le i<j\le m,
\end{equation}
and
\begin{equation}
\label{hatQ-proj}
\hatQ_{\theta_{\bt_1}(\AA_1)}\dots\hatQ_{\theta_{\bt_m}(\AA_m)}=\hatQ_{\theta_{\bt_1}(\AA_1)\,\cap\dots\cap\,\theta_{\bt_m}(\AA_m)}.
\end{equation}
Finally, $\hatQ$ of the full space (i.e., $(\scrS_2)^{|\T_L|}$) is the unity, $\hatQ_\emptyset=0$, and if~$\AA_1,\AA_2,\dots$ is a countable collection of disjoint events, then (in the strong-operator topology)
\begin{equation}
\label{QSigmaAdd}
\hatQ_{\bigcup_{n=1}^\infty \AA_n}=\sum_{n=1}^\infty \hatQ_{\AA_n}.
\end{equation} 
In particular,~$\hatQ_{\AA^\cc}=\1-\hatQ_{\AA}$ for any event~$\AA$.
\end{myproposition}

\begin{proofsect}{Proof}
The map~$\whvartheta_{\bt}$ is a pure reflection for even-parity~$\bt\in\T_{L/B}$ and so \eqref{Q-theta} holds by the fact that pure reflection of~$\hatQ_{\AA}$ is $\hatQ$ of the reflected~$\AA$. For odd-parity~$\bt$, the relation \eqref{conjugate} implies $\overline{\hatQ_\AA}=\hatQ_{\sigma(\AA)}$, which yields \eqref{Q-theta} in these cases as well.
The remaining identities are easy consequences of the definitions and \eqref{resolution}.
\end{proofsect}

\begin{myremark}
\label{POV-measure}
The last few properties listed in the lemma imply that the map $\AA \to \hatQ_{\AA}$ is a \emph{positive-operator-valued (POV) measure}, in the sense of \cite{Davies}. As a consequence, if $\AA \subset \AA'$ then $\hatQ_{\AA} \leq \hatQ_{\AA'}$ while if $\{\AA_n\}$ is a countable collection of events, not necessarily disjoint, then
\begin{equation}
\label{Q-subadditive}
\hatQ_{\bigcup_{n=1}^\infty \AA_n}\le \sum_{n=1}^\infty \hat{Q}_{\AA_n}.
\end{equation}
Both of these properties are manifestly true by the definition \eqref{hatQ}.
\end{myremark}

Before we state our next theorem, let us recall the ``standard'' setting for the application of chessboard estimates to proofs of phase transitions in \emph{classical} models. Given~$B$ that divides~$L$, one typically singles out a collection~$\GG_1,\dots,\GG_n$ of ``good'' $B$-block events and defines
\begin{equation}
\label{bad-event}
\BB=(\GG_1\cup\dots\cup\GG_n)^\cc
\end{equation}
to be the corresponding ``bad'' $B$-block event. Without much loss of generality we will assume that~$\BB$ is invariant under ``complex'' reflections, i.e.,~$\vartheta_{\bt}(\BB)=\tau_{B\bt}(\BB)$, where~$\tau_{\br}$ denotes the shift by~$\br$ on~$(\scrS_2)^{|\T_L|}$. 
In the best of situations, carefully chosen good events typically satisfy the conditions in the following definition:

\begin{mydefinition}
\label{def-incompatibility}
We say that the ``good'' $B$-block events are \emph{incompatible} if
\settowidth{\leftmargini}{(11)}
\begin{enumerate}
\item[(1)]
they are mutually exclusive, i.e., $\GG_i\cap\GG_j=\emptyset$ whenever~$i\ne j$;
\item[(2)]
their simultaneous occurrence at neighboring blocks forces an intermediate block (which overlaps the two neighbors) i.e., there exists~$\ell$ with~$1\le\ell<B$ such that
\begin{equation}
\label{incompatibility}
\theta_{\bt}(\GG_i)\cap\theta_{\bt'}(\GG_j)\subset\tau_{B\bt+\ell(\bt'-\bt)}(\BB)
\end{equation}
holds for all~$i\ne j$ and any~$\bt,\bt'\in\T_{L/B}$ with~$|\bt-\bt'|=1$. Here~$\tau_{\br}$ is the shift by~$\br$.
\end{enumerate}
\end{mydefinition}

These conditions are much easier to achieve in situations where we are allowed to use reflections through planes containing sites. Then, typically, one defines the~$\GG_i$'s so that the neighboring blocks \emph{cannot} have distinct types of goodness. But as noted in Remark~\ref{no-quantum-CE}, we are not allowed to use these reflections in the quantum setting. Nevertheless,~(1) and~(2) taken together do ensure that a simultaneous occurrence of two distinct types of goodness necessarily enforces a ``contour'' of bad blocks. The weight of each such contour can be bounded by the quantity~$\frakp_{L,\beta}(\BB)$ to the number of constituting blocks; it then remains to show that $\frakp_{L,\beta}(\BB)$ is sufficiently small. For quantum models, appropriate modifications of this strategy yield the following result:

\begin{theorem}
\label{thm-QPT}
Consider a quantum spin system on~$\T_L$ with spin~$\CalS$ and interaction for which the Gibbs state~$\state{-}_{L,\beta}$ from~\eqref{thermal-state} is reflection positive for reflections through planes between sites on~$\T_L$. Let~$H^\infty$ be a function and $\xi>0$ a constant such that, for all~$L\ge1$,
\begin{equation}
\label{xi-def}
\sup_{\Omega\in(\scrS_2)^{|\T_L|}}\bigl|[H]_\Omega-H^\infty(\Omega)\bigr|
+
\sup_{\Omega\in(\scrS_2)^{|\T_L|}}\bigl|\state{H}_\Omega-H^\infty(\Omega)\bigr|
\le\xi\,|\T_L|.
\end{equation}
Let~$\GG_1,\dots,\GG_n$ be incompatible ``good'' $B$-block events and define~$\BB$ as in \eqref{bad-event}.
Suppose that~$\BB$ is invariant under reflections and conjugation~$\sigma$, i.e., $\vartheta_{\bt}(\BB)=\tau_{B\bt}(\BB)$ for all~$\bt\in\T_{L/B}$.
Fix~$\epsilon>0$. Then there exists~$\delta>0$ such that if~$\beta\le c_2\sqrt\CalS$ and
\begin{equation}
\label{class-cond}
\frakp_{L,\beta}(\BB)\,\texte^{\,\beta(\xi+c_3/\sqrt{\CalS})}<\delta,
\end{equation}
where~$c_2$ and $c_3$ are as in Theorem~\ref{thm-matrix-element}, we have
\begin{equation}
\label{bad-bad}
\bigl\langle\hatQ_{\BB}\bigr\rangle_{L,\beta}<\epsilon
\end{equation}
and, for all~$i=1,\dots,n$ and all distinct~$\bt_1,\bt_2\in\T_{L/B}$,
\begin{equation}
\label{no-clustering}
\state{\hatQ_{\theta_{\bt_1}(\GG_i)}[\1-\hatQ_{\theta_{\bt_2}(\GG_i)}]}_{L,\beta}<\epsilon.
\end{equation}
Here~$\delta$ may depend on~$\epsilon$ and~$d$, but not on~$\beta$, $\CalS$, $n$ nor on the details of the model.
\end{theorem}

\begin{myremarks}
Here are some notes concerning the previous theorem:
\settowidth{\leftmargini}{(11)}
\begin{enumerate}
\item[(1)]
By general results (e.g.,~\cite{Duffield}) on the convergence of upper and lower symbols as~$\CalS\to\infty$, the quantity~$\xi$ in \eqref{xi-def} can be made arbitrarily small by increasing~$\CalS$ appropriately. In fact, for two-body interactions,~$\xi$ is typically a small constant times~$1/\CalS$ and so it provides a harmless correction to the term~$c_3/\sqrt\CalS$ in \eqref{class-cond}. In particular, apart from the classical bound that~$\frakp_{L,\beta}(\BB)\ll1$, \eqref{class-cond} will only require that~$\beta\ll\sqrt\CalS$.
\item[(2)]
Note that the result is stated for pure reflections, $\theta_{\bt}(\GG_i)$, of the good events, not their more complicated counterparts~$\vartheta_{\bt}(\GG_i)$. This is important for maintaining a close link between the nature of phase transition in the quantum model and its classical counterpart. We also note that~$H^\infty$ is \emph{not} required to be reflection positive for Theorem~\ref{thm-QPT} to hold. (Notwithstanding, the classical Hamiltonian will be reflection positive for all examples in Sect.~\ref{sec5}.)
\item[(3)]
The stipulation that the~$\vartheta_{\bt}$'s ``act'' on~$\BB$ only as translations is only mildly restrictive: Indeed,~$\sigma(\BB)=\BB$ in all cases treated in the present work. However, if it turns out that ~$\sigma(\BB)\ne\BB$, the condition \eqref{class-cond} may be replaced by
\begin{equation}
\label{class-cond2}
\sqrt{\frakp_{L,\beta}(\BB)\frakp_{L,\beta}\bigl(\sigma(\BB)\bigr)}\,\texte^{\,\beta(\xi+c_3/\sqrt{\CalS})}<\delta,
\end{equation}
which---since $\frakp_{L,\beta}(\sigma(\BB))\le1$---is anyway satisfied by a stricter version of \eqref{class-cond} (this does need reflection positivity of~$H^\infty$). Note that~$\sigma(\BB)=\BB$ implies that every configuration in~$\sigma(\GG_i)$ is also good. In most circumstances we expect that~$\sigma(\GG_i)$ is one of the good events.
\end{enumerate}
\end{myremarks}

\subsection{Phase transitions in quantum models}
\label{sec3.3}\noindent
It remains to show how to adapt the main conclusion of Theorem~\ref{thm-QPT} to the proof of phase transition in quantum systems. We first note that \eqref{class-cond} is a condition on the \emph{classical} model which, for~$\delta$ small, yields a classical variant of \eqref{no-clustering},
\begin{equation}
\label{no-clustering-class}
\BbbP_{L,\beta}\bigl(\theta_{\bt_1}(\GG_i)\cap\theta_{\bt_2}(\GG_i^\cc)\bigr)<\epsilon, \qquad 1\le i\le n.
\end{equation}
Under proper conditions on~$\epsilon$ and the probabilities of the~$\GG_i$'s, this yields absence of clustering for the classical torus Gibbs state which, by a conditioning ``on the back of the torus''---see the paragraph before Lemma~\ref{lemma-KMS}---implies the existence of multiple infinite-volume Gibbs measures.

For a quantum system with an internal symmetry, a similar argument allows us to deal with the cases when the symmetry has been ``spontaneously'' broken. For instance (see~\cite{FL}) in magnetic systems \eqref{no-clustering} might imply the non-vanishing of the spontaneous magnetization which, in turn, yields a discontinuity in some derivative of the free energy, i.e., a \emph{thermodynamic} phase transition. In the cases with no symmetry---or in situations where the symmetry is not particularly useful, such as for temperature-driven phase transitions---we can still demonstrate a thermodynamic transition either by concocting an ``unusual'' external field (which couples to distinct types of good blocks) or by directly proving a jump e.g.\ in the energy density. 

An elegant route to these matters is via the formalism of infinite-volume KMS states (see, e.g., \cite{Israel,Simon-book}). Let us recall the principal aspects of this theory: Consider the
$C^\star$~algebra~$\frakA$ of quasilocal observables defined as the norm-closure of $\bigcup_{\Lambda\subset \Z^d} \frakA_{\Lambda}$, where the union is over all finite subsets $\Lambda$ and where $\frakA_\Lambda$ is the set of all bounded operators on the Hilbert space~$\HH_{\Lambda}=\bigotimes_{\br\in\Lambda}[\C^{2\CalS+1}]_{\br}$. (To interpret the union properly, we note that if $\Lambda\subset\Lambda'$, then $\frakA_{\Lambda}$ is isomorphic to a subset of~$\frakA_{\Lambda'}$, via the map~$A\to A\otimes\1$ with~$\1$ being the identity in $\frakA_{\Lambda'\setminus\Lambda}$.) For each~$L\ge1$, let us identify~$\T_L$ with the block~$\Lambda_L$ and let~$H_L$ be the Hamiltonian on~$\T_L$ which we assume is of the form~\eqref{Hamiltonian} with~$h_\Gamma$ finite range and translation invariant. 

For each observable~$A\in\frakA_{\Lambda_L}$, let~$\alpha^{(L)}_t(A)=\texte^{\texti tH_L}A\texte^{-\texti tH_L}$ be the strongly-conti\-nu\-ous one-parameter family of operators representing the time evolution of~$A$ in the Heisenberg picture. For~$A$ local and~$H_L$ finite range, by expanding into a series of commutators
\begin{equation}
\label{alphat}
\alpha_t^{(L)}(A)=\sum_{n\ge0}\frac{(\texti t)^n}{n!}[H_L[H_L\dots[H_L,A]\dots]],
\end{equation}
the map $t\mapsto\alpha^{(L)}_t(A)$ extends to all~$t\in\C$, see~\cite[Theorem~III.3.6]{Israel}. Moreover, the infinite series representation of $\alpha_t^{(L)}(A)$ converges in norm, as~$L\to\infty$, to a one-parameter family of operators~$\alpha_t(A)$, uniformly in~$t$ on compact subsets of~$\C$. (These facts were originally proved in \cite{Robinson}.) 

A state~$\langle-\rangle_\beta$ on~$\frakA$---i.e., a linear functional obeying~$\state{A}_\beta\ge0$ if~$A\ge0$ and $\state{\1}_\beta=1$---is called a \emph{KMS state} (for the translation-invariant, finite-range interaction~$H$ at inverse temperature~$\beta$) if for all local operators~$A,B\in\frakA$, the equality
\begin{equation}
\label{KMS}
\state{AB}_\beta=\state{\alpha_{-\texti\beta}(B)A}_\beta,
\end{equation}
also known as the \emph{KMS condition}, holds. This condition is the quantum counterpart of the DLR equation from classical statistical mechanics and a KMS state is thus the counterpart of the infinite-volume Gibbs measure.

We proceed by stating two general propositions which will help us apply the results from previous sections to the proof of phase transitions. We begin with a statement which concerns phase transitions due to symmetry breaking:

\begin{myproposition}
\label{prop-SB}
Consider the quantum spin systems as in Theorem~\ref{thm-QPT} and suppose that the incompatible good block events $\GG_1,\dots,\GG_n$ are such that~$\langle \hatQ_{\GG_k}\rangle_{L,\beta}$ is the same for all~$k=1,\dots,n$. If~\twoeqref{bad-bad}{no-clustering} hold with an~$\epsilon$ such that~$(n+1)\epsilon<\ffrac12$, then there exist~$n$ distinct, KMS states~$\state{-}_\beta^{(k)}$, $k=1,\dots,n$, which are invariant under translations by~$B$ and for which
\begin{equation}
\label{3.29}
\bigl\langle\hatQ_{\GG_k}\bigr\rangle_\beta^{(k)}\ge1-(n+1)\epsilon,
\qquad k=1,\dots,n.
\end{equation}
\end{myproposition}

The proposition says that there are at least $n$ distinct equilibrium states. There may be more, but not less. This ensures
a phase transition, via phase coexistence.

Our second proposition deals with temperature driven transitions. The following is a quantum version of one of the principal theorems in~\cite{Kotecky-Shlosman,KS-proceedings}:

\begin{myproposition}
\label{prop-EE}
Consider the quantum spin systems as in Theorem~\ref{thm-QPT} and let~$\GG_1$ and~$\GG_2$ be two incompatible $B$-block events. Let~$\beta_1<\beta_2$ be two inverse temperatures and suppose that~$\epsilon\in[0,\ffrac14)$ is such that for all~$L\ge1$,
\settowidth{\leftmargini}{(1111)}
\begin{enumerate}
\item[(1)]
the bounds \twoeqref{bad-bad}{no-clustering} hold for all~$\beta\in[\beta_1,\beta_2]$,
\item[(2)]
$\langle\hatQ_{\GG_1}\rangle_{L,\beta_1}\ge1-2\epsilon$ and $\langle\hatQ_{\GG_2}\rangle_{L,\beta_2}\ge1-2\epsilon$.
\end{enumerate}
Then there exists an inverse temperature~$\betat\in[\beta_1,\beta_2]$ and two distinct KMS states $\state{-}_{\betat}^{(1)}$ and~$\state{-}_{\betat}^{(2)}$ at inverse temperature~$\betat$ which are invariant under translations by~$B$ and for which
\begin{equation}
\bigl\langle\hatQ_{\GG_1}\bigr\rangle_{\betat}^{(1)}\ge1-4\epsilon
\quad\text{and}\quad
\bigl\langle\hatQ_{\GG_2}\bigr\rangle_{\betat}^{(2)}\ge1-4\epsilon.
\end{equation}
\end{myproposition}

The underlying idea of the latter proposition is the existence of a forbidden gap in the density of, 
say,~$\GG_1$-blocks. Such ``forbidden gap'' arguments have been invoked in (limiting) toroidal states by, e.g.,~\cite{Kotecky-Shlosman,KS-proceedings,Gawedzki}; an extension to infinite-volume, translation-invariant, reflection-positive  Gibbs states has appeared in~\cite{BK}. Both propositions are proved in Sect.~\ref{sec4.3}.

\section{Proofs}
\label{sec4}\noindent
Here we provide the proofs of our general results from Sect.~\ref{sec3}. We begin by the estimates of matrix elements of Gibbs-Boltzmann weight (Theorem~\ref{thm-matrix-element}) and then, in Sect.~\ref{sec4.2}, proceed to apply these in quasiclassical Peierls' arguments which lie at the core of Theorem~\ref{thm-QPT}. Finally, in Sect.~\ref{sec4.3}, we elevate the conclusions of Theorem~\ref{thm-QPT} to coexistence of multiple KMS states, thus proving Propositions~\ref{prop-SB}-\ref{prop-EE}.

\subsection{Bounds on matrix elements}
\noindent
The proof of Theorem~\ref{thm-matrix-element} is based on a continuity argument whose principal estimate is encapsulated into the following claim:

\begin{myproposition}
\label{prop4.1}
Suppose that \twoeqref{range}{Lipschitz} hold with constants~$R$, $c_0$, and~$c_1$. Let~$\wh H_\Omega=H-[H]_\Omega$. Suppose there exist~$c_2>0$ and~$\epsilon>0$ such that for all $\beta\le c_2\sqrt\CalS$, 
\begin{equation}
\label{element2}
\bigl|\langle\Omega|\texte^{-\beta\wh H_\Omega}|\Omega'\rangle\bigr|
\le\texte^{-\eta\dall{\Omega}{\Omega'}+\beta\epsilon|\Lambda|}
\end{equation}
is true for all~$\Omega,\Omega'\in(\scrS_2)^{|\Lambda|}$.
Then there exists a constant~$c_3$ depending on $c_0$, $c_1$, $c_2$ and $R$ (but not $\Lambda$, $\CalS$ or $\epsilon$) such that for all $\beta\le c_2\sqrt\CalS$, 
\begin{equation}
\label{derivative-bd}
\Bigl|\frac\textd{\textd\beta}\langle\Omega|\texte^{-\beta\wh H_\Omega}|\Omega'\rangle\Bigr|
\le\frac{c_3}{\sqrt{\CalS}}\,|\Lambda|\,\texte^{-\eta\dall\Omega{\Omega'}+\beta\epsilon|\Lambda|}.
\end{equation}
\end{myproposition}

Before we commence with the proof, we will make a simple observation:

\begin{mylemma}
\label{lemma-triangle}
For all~$\Lambda$ and all~$\Omega,\Omega',\Omega''\in(\scrS_2)^{|\Lambda|}$,
\begin{equation}
\dall\Omega{\Omega'}\le\dall{\Omega'}{\Omega''}+\done\Omega{\Omega''}
+\sum_{\br\in\Lambda}1_{\{\Omega_{\br}\ne\Omega''_{\br}\}}.
\end{equation}
\end{mylemma}

\begin{proofsect}{Proof}
Since all ``norms'' in the formula are sums over~$\br\in\Lambda$, it suffices to prove the above for~$\Lambda$ having only one point. This is easy: For $\Omega=\Omega''$ the inequality is actually an equality. Otherwise, we apply the bounds $\dall\Omega{\Omega'}\le\sqrt\CalS|\Omega-\Omega'|$ and $\dall{\Omega'}{\Omega''}+1\ge\sqrt\CalS|\Omega'-\Omega''|$ to convert the statement into the triangle inequality for the $\ell^1$-norm.
\end{proofsect}

\begin{proofsect}{Proof of Proposition~\ref{prop4.1}}
Let us fix~$\Omega$ and~$\Omega'$ for the duration of this proof and abbreviate~$M(\beta)=\langle\Omega|\texte^{-\beta\wh H_\Omega}|\Omega'\rangle$.
We begin by expressing the derivative of $M(\beta)$ as an integral over coherent states.
Indeed, $M'(\beta)=-\langle\Omega|\wh H_\Omega\,\texte^{-\beta \wh H_\Omega}|\Omega'\rangle$ and so inserting the upper-symbol representation~\eqref{upper-symbol} for~$\wh H_\Omega=\sum_{\Gamma\subset\Lambda}(h_\Gamma-[h_\Gamma]_\Omega)$, we have
\begin{equation}
\label{3.8}
M'(\beta)
=-\sum_{\Gamma\subset\Lambda}\left(\frac{2\CalS+1}{4\pi}\right)^{|\Lambda|}
\int_{(\scrS_2)^{|\Lambda|}}\!\!\textd\wt\Omega''\,\,\langle\Omega|\wt\Omega''\rangle
\langle\wt\Omega''|\texte^{-\beta\wh H_\Omega}|\Omega'\rangle
\bigl([h_\Gamma]_{\wt\Omega''}-[h_\Gamma]_\Omega\bigr).
\end{equation}
By the fact that~$[h_\Gamma]_{\wt\Omega''}-[h_\Gamma]_\Omega$ depends only on the portion of~$\wt\Omega''$ on~$\Gamma$, the integrals over the components of~$\wt\Omega''$ outside~$\Gamma$ can be carried out which yields
\begin{equation}
\label{3.9}
M'(\beta)
=-\!\sum_{\Gamma\subset\Lambda}\left(\frac{2\CalS+1}{4\pi}\right)^{|\Gamma|}\!\!
\int_{(\scrS_2)^{|\Gamma|}}\!\!\textd\Omega_\Gamma''\,\langle\Omega_\Gamma|\Omega_\Gamma''\rangle
\langle\Omega''|\texte^{-\beta\wh H_\Omega}|\Omega'\rangle
\bigl([h_\Gamma]_{\Omega''}-[h_\Gamma]_\Omega\bigr).
\end{equation}
Here, as for the rest of this proof,~$\Omega''$ is set to~$\Omega$ outside~$\Gamma$ and to~$\Omega_\Gamma''$ in~$\Gamma$.

Let~$\II_\Gamma$ denote the integral on the right-hand side of~\eqref{3.9}. Using~\eqref{d-eta},~\eqref{element2} and~\eqref{Lipschitz} we have
\begin{equation}
\label{1st-bd}
|\II_\Gamma|\le c_1\Vert h_\Gamma\Vert\,\texte^{\beta\epsilon|\Lambda|}
\!\!\int_{(\scrS_2)^{|\Gamma|}}\!\!\textd\Omega_\Gamma''\,
\texte^{-\eta\dall{\Omega'}{\Omega''}-\eta\dtwo{\Omega''}{\Omega}-\beta([H]_{\Omega''}-[H]_\Omega)}\norm{\Omega''}{\Omega}_1.
\end{equation}
(Recall from the definition that 
 $\wh H_{\Omega} = \wh H_{\Omega''} - [H]_{\Omega} + [H]_{\Omega''}$.)
In order to bound the right-hand side, we need a few simple estimates. First, noting that
\begin{equation}
[H]_{\Omega''} - [H]_{\Omega}\,=\sum_{\Gamma':\Gamma'\cap\Gamma\neq \emptyset} ([h_{\Gamma'}]_{\Omega''} - [h_{\Gamma'}]_{\Omega})\, ,
\end{equation}
\eqref{uniform-bd}~and~\eqref{Lipschitz} imply that, for some constant~$c_4$ depending only on~$c_0$,~$c_1$ and $R$,
\begin{equation}
\bigl|[H]_{\Omega''}-[H]_\Omega\bigr|\le c_4\norm{\Omega}{\Omega}_1=c_4\norm{\Omega_\Gamma''}{\Omega_\Gamma}_1.
\end{equation} 
Second, Lemma~\ref{lemma-triangle} tells us
\begin{equation}
-\dall{\Omega'}{\Omega''}\le-\dall{\Omega}{\Omega'}+\done{\Omega_\Gamma}{\Omega_\Gamma''}+|\Gamma|.
\end{equation}
Finally, $\norm{\Omega''}{\Omega}_1$ is bounded by~$\CalS^{-1/2}$ times the exponential of~$\done{\Omega}{\Omega''}$. Since we are assuming that~$\beta\le c_2\sqrt\CalS$, we conclude that
\begin{equation}
\texte^{-\eta\dall{\Omega'}{\Omega''}-\beta([H]_{\Omega''}-[H]_\Omega)}\norm{\Omega''}{\Omega}_1
\le\frac{\texte^{\eta|\Gamma|}}{\sqrt{\CalS}}\,\texte^{-\eta\dall{\Omega}{\Omega'}+c_5\done{\Omega_\Gamma}{\Omega_\Gamma''}}
\end{equation}
for some constant~$c_5$ independent of~$\CalS$ and~$\Lambda$.

Plugging this back in the integral \eqref{1st-bd},  we get
\begin{equation}
\label{IGammabd}
|\II_\Gamma|\le \frac{c_1\texte^{\eta|\Gamma|}}{\sqrt{\CalS}}\,\Vert h_\Gamma\Vert\,\,\texte^{\beta\epsilon|\Lambda|-\eta\dall\Omega{\Omega'}}
\int_{(\scrS_2)^{|\Gamma|}}\textd\Omega_\Gamma''\,\,
\texte^{c_5\done{\Omega_\Gamma}{\Omega_\Gamma''}-\eta\dtwo{\Omega_\Gamma}{\Omega_\Gamma''}}.
\end{equation}
To estimate the integral, we note that both norms in the exponent are sums over individual components. Hence, the integral is bounded by the product of~$|\Gamma|$ integrals of the form
\begin{equation}
K=\int_{\scrS_2}\textd\br''\,\,\texte^{c_5\sqrt\CalS|\br-\br''|-\eta \CalS|\br-\br''|^2},
\end{equation}
where~$\br$ and~$\br''$ are vectors on~$\scrS_2$---representing the corresponding 3-dimensional components of~$\Omega_\Gamma$ and~$\Omega_\Gamma''$---and where $|\br-\br''|$  denotes Euclidean distance in~$\R^3$. 
Parametrizing by~$r=|\br-\br''|$ and integrating over the polar angle of~$\br''$ relative to~$\br$, we now get
\begin{equation}
\label{4.11}
K= \int_0^2\textd r\scrJ(r)\,\texte^{-\frac12\eta S\, r^2+c_5\sqrt{\CalS}\,r}.
\end{equation}
Here the Jacobian,~$\scrJ(r)$, is the circumference of the circle~$\{\br''\colon|\br''|=1,\,|\br-\br''|=r\}$. But this circle has radius smaller than~$r$ and so~$\scrJ(r)\le2\pi r$. Scaling~$r$ by~$\CalS^{-1/2}$ yields $K\le c_6/\CalS$ for some constant~$c_6>0$ independent of~$\CalS$. 

Plugging this back in~\eqref{IGammabd}, we then get
\begin{equation}
\label{last-bd}
|\II_\Gamma|\le
\frac{c_1}{\sqrt{\CalS}}\Bigl(\frac{c_6 \texte^{\eta}}\CalS\Bigr)^{|\Gamma|}\Vert h_\Gamma\Vert
\,\texte^{-\eta\dall\Omega{\Omega'}+\beta\epsilon|\Lambda|}.
\end{equation}
Inserting this into~\eqref{3.9}, using~\eqref{range} to bound the terms exponential in~$|\Gamma|$ by a constant depending only on~$R$---this is possible because there are~$|\Gamma|$ factors of~$\CalS$'s in the denominator of~\eqref{last-bd} that can be used to cancel the factors~$(2\CalS+1)$ in front of the integral in~\eqref{3.9}---and applying~\eqref{uniform-bd}, we get~\eqref{derivative-bd}.
\end{proofsect}

On the basis of Proposition~\ref{prop4.1}, the proof of Theorem~\ref{thm-matrix-element} is easily concluded:

\begin{proofsect}{Proof of Theorem~\ref{thm-matrix-element}}
Let~$c_2$ and~$c_3$ be the constants from Proposition~\ref{prop4.1} and let~$\epsilon=c_3/\sqrt\CalS$. We claim that \eqref{element2} holds for all~$\beta\le c_2\sqrt\CalS$. First, in light of \eqref{d-eta} and the definition of~$\dall{\Omega}{\Omega'}$, \eqref{element2} holds for~$\beta=0$. This allows us to define~$\beta_0$ to be the largest number such that \eqref{element2} holds for all~$\beta\in[0,\beta_0]$. Now, if~$\beta\le \beta_0\wedge c_2\sqrt\CalS$, then Proposition~\ref{prop4.1} and our choice of~$\epsilon$ guarantee that the~$\beta$-derivative of $\langle\Omega|\texte^{-\beta\wh H_\Omega}|\Omega'\rangle$ is no larger than that of the right-hand side of \eqref{derivative-bd}. We deduce (by continuity) that~$\beta_0=c_2\sqrt\CalS$. Using that~$\wh H_\Omega=H-[H]_\Omega$, we now get \eqref{element}.
\end{proofsect}

\begin{proofsect}{Proof of Corollary~\ref{cor3.2}}
First we observe that the diagonal matrix element $\langle\Omega|\texte^{-\beta H}|\Omega\rangle$ is real and positive. The upper bound is then the~$\Omega'=\Omega$ version of Theorem~\ref{thm-matrix-element}; the lower bound is a simple consequence of Jensen's---also known as the Peierls-Bogoliubov---inequality; see, e.g.,~\cite[Theorem~I.4.1]{Simon-book}.
\end{proofsect}

\subsection{Quasiclassical Peierls' arguments}
\label{sec4.2}\noindent
Our goal is to prove the bounds \twoeqref{bad-bad}{no-clustering}. To this end, let us introduce the quantum version of the quantity from \eqref{frakp}: For any~$B$-block event~$\AA$, let
\begin{equation}
\frakq_{L,\beta}(\AA)=\state{\,\prod_{\bt\in\T_{L/B}}\hatQ_{\vartheta_{\bt}(\AA)}}_{L,\beta}^{(B/L)^d}.
\end{equation}
(Note that, by \eqref{Q-theta}, this is of the form of the expectation on the right hand side of~\eqref{chessboard}.)
First we will note the following simple consequence of Theorem~\ref{thm-matrix-element}:

\begin{mylemma}
\label{lemma-qp}
Let~$\xi$ be as in \eqref{xi-def} and let~$c_2$ and~$c_3$ be as in Theorem~\ref{thm-matrix-element}. If~$\beta\le c_2\sqrt{\CalS}$, then for any $B$-block event~$\AA$,
\begin{equation}
\label{qp-rel}
\frakq_{L,\beta}(\AA)\le\bigl[\frakp_{L,\beta}(\AA)\frakp_{L,\beta}\bigl(\sigma(\AA)\bigr)\bigr]^{\ffrac12}
\,\,\texte^{\,\beta(\xi+c_3/\sqrt\CalS)}.
\end{equation}
\end{mylemma}

\begin{proofsect}{Proof}
By~\eqref{hatQ-proj} we have
\begin{equation}
\frakq_{L,\beta}(\AA)=\langle\hatQ_{\wt\AA}\rangle_{L,\beta}^{(B/L)^d}
\quad\text{where}\quad
\wt\AA=\bigcap_{\bt\in\T_{L/B}}\vartheta_{\bt}(\AA).
\end{equation}
Invoking the integral representation \eqref{hatQ}, the bounds from Corollary~\ref{cor3.2} and the definition of~$\xi$ from~\eqref{xi-def},
\begin{equation}
\frakq_{L,\beta}(\AA)\le\BbbP_{L,\beta}(\wt\AA)^{(B/L)^d}\texte^{\,\beta(\xi+c_3/\sqrt\CalS)}.
\end{equation}
Now we may use \eqref{class-chessboard} for the classical probability and we get \eqref{qp-rel}.
\end{proofsect}

Next we will invoke the strategy of~\cite{FL} to write a bound on the correlator in \eqref{no-clustering} in terms of a sum over Peierls contours. Let~$\scrM_{L/B}$ denote the set of connected sets~$\Y \subset\T_{L/B}$ with connected complement. By a \emph{contour} we then mean the boundary of a set~$\Y\in\scrM_{L/B}$, i.e., the set~$\partial\Y$ of nearest neighbor edges on~$\T_{L/B}$ with one endpoint in~$\Y$ and the other endpoint in~$\Y^\cc\subset\T_{L/B}$. The desired bound is as follows:

\begin{mylemma}
Let~$\GG_1,\dots,\GG_n$ be incompatible good events and let~$\BB$ be the bad event with the property that $\tau_{B\bt}(\BB)=\vartheta_{\bt}(\BB)$ for all~$\bt\in\T_{L/B}$.
Then for all distinct~$\bt_1,\bt_2\in\T_{L/B}$ and all~$i=1,\dots,n$,
\begin{equation}
\label{sum-Lambda}
\state{\hatQ_{\theta_{\bt_1}(\GG_i)}\hatQ_{\theta_{\bt_2}(\GG_i^\cc)}}_{L,\beta}
\le\,\, \!\!\!\sum_{\begin{subarray}{c} \Y
\colon \Y \in\scrM_{L/B}\\\bt_1\in\Y,\,\bt_2\not\in\Y
\end{subarray}}
2\,\bigl[4\frakq_{L,\beta}(\BB)\bigr]^{\frac1{4d}|\partial\Y|}.
\end{equation}
\end{mylemma}

\begin{proofsect}{Proof}
We begin by noting that $\bt_1\ne\bt_2$ and \twoeqref{hatQ-commute}{hatQ-proj} give us
\begin{equation}
\label{QQ-integral}
\hatQ_{\theta_{\bt_1}(\GG_i)}\hatQ_{\theta_{\bt_2}(\GG_i^\cc)}
=\left(\frac{2\CalS+1}{4\pi}\right)^{|\T_L|}
\int_{\theta_{\bt_1}(\GG_i)\cap\theta_{\bt_2}(\GG_i^\cc)}
\!\!\!\!\!\textd\Omega\,\,\proj\Omega.
\end{equation}
Now pick~$\Omega\in\theta_{\bt_1}(\GG_i)\cap\theta_{\bt_2}(\GG_i^\cc)$ and let~$\Y'\subset\T_{L/B}$ be  the largest connected component of~$B$-blocks---i.e., translates of~$\Lambda_B$ by~$B\bt$, with~$\bt\in\T_{L/B}$---such that~$\bt_1\in\Y'$ and that~$\theta_{\bt}(\GG_i)$ occurs for every~$\bt\in\Y'$.
This set may not have connected complement, so we define~$\Y\in\scrM_{L/B}$ to be the set obtained by filling the ``holes'' of~$\Y'$, \emph{except} that which contains~$\bt_2$. Note that all translates of~$\Lambda_B$ corresponding to the boundary sites of~$\Y$ are of type~$\GG_i$. 

In order to extract the weight of the contour, we will have to introduce some more notation. Decomposing the set of boundary edges $\partial\Y$ into~$d$ sets
$\partial_1\Y,\dots,\partial_d\Y$ according to the coordinate directions 
into which the edges are pointing, let~$j$ be a direction where~$|\partial_j\Y|$ is maximal. Furthermore, let~$\Y_j^{\text{ext}}$ be the set of sites in~$\Y^\cc$ which are on the ``left'' side of an edge in~$\partial_j\Y$. It is easy to see that this singles out exactly half of the sites in~$\Y^\cc$ that are at the endpoint of an edge in~$\partial_j\Y$.
Next we intend to show that the above setting implies the existence of at least~$|\Y_j^{\text{ext}}|/2$ bad blocks whose position is more or less determined by~$\Y$.

Recall that~$\hate_j$ denotes the unit vector in the~$j$-th coordinate direction. Since the good events satisfy the incompatibility condition \eqref{incompatibility}, at least one of the following two possibilities must occur: either $\Omega\in\tau_{B\bt}(\BB)$ for at least half of~$\bt\in\Y_j^{\text{ext}}$ or $\Omega\in\tau_{B\bt+\ell\hate_j}(\BB)$ for at least half of~$\bt\in\Y_j^{\text{ext}}$. (Here~$\ell$ is the constant from the definition of incompatibility.) Indeed, if the former does not occur then more than half of~$\bt\in\Y_j^{\text{ext}}$ mark a good block, but of a different type of goodness than~$\GG_i$. Since this block neighbors on a~$\GG_i$-block, incompatibility of good block events implies that a bad block must occur~$\ell$ lattice units along the line between these blocks.

Let us temporarily abbreviate~$K_j=|\Y_j^{\text{ext}}|$ and let~$\scrC_j(\Y)$ be the set of collections of~$K_j/2$ sites representing the positions of the aforementioned $K_j/2$ bad blocks. In light of $\tau_{B\bt}(\BB)=\vartheta_{\bt}(\BB)$, the above argument implies
\begin{equation}
\label{ContourDecomp}
\theta_{\bt_1}(\GG_i) \cap \theta_{\bt_2}(\GG_i^\cc)
\subset\,\, \!\!\!\bigcup_{\begin{subarray}{c}
\Y\colon \Y\in\scrM_{L/B}\\\bt_1\in\Y,\,\bt_2\not\in\Y
\end{subarray}}\,\, 
\bigcup_{(\bt_i)\in\scrC_j(\Y)}\left(\,
\bigcap_{i=1}^{K_j/2}\bigl(\vartheta_{\bt_i}(\BB)\bigr) \cup \bigcap_{i=1}^{K_j/2}\tau_{\ell\hate_j}\bigl(\vartheta_{\bt_i}(\BB)\bigr)\right).
\end{equation}
Therefore, using the fact that $\AA\mapsto\hatQ_\AA$ is a POV measure (cf Remark~\ref{POV-measure}), this implies
\begin{equation}
\label{4.21eq}
\hatQ_{\theta_{\bt_1}(\GG_i)}\hatQ_{\theta_{\bt_2}(\GG_i^\cc)}
\le\,\, \!\!\!\sum_{\begin{subarray}{c}
\Y\colon \Y\in\scrM_{L/B}\\\bt_1\in\Y,\,\bt_2\not\in\Y
\end{subarray}}\,\,
\sum_{(\bt_i)\in\scrC_j(\Y)}\left(\,
\prod_{i=1}^{K_j/2}\hatQ_{\vartheta_{\bt_i}(\BB)}+\prod_{i=1}^{K_j/2}\hatQ_{\tau_{\ell\hate_j}\bigl(\vartheta_{\bt_i}(\BB)\bigr)}\right).
\end{equation}
Here the two terms account for the two choices of where the bad events can occur and~$j$ is the direction with maximal projection of the boundary of~$\Y$ as defined above. Since \eqref{chessboard}, \eqref{Q-theta} and~$\theta(\BB)=\BB$ allow us to conclude that
\begin{equation}
\state{\,\prod_{i=1}^{K_j/2}\hatQ_{\vartheta_{\bt_i}(\BB)}}_{L,\beta}
\le\frakq_{L,\beta}(\BB)^{K_j/2},
\end{equation}
and since the translation invariance of the torus state~$\state{-}_{L,\beta}$ implies a similar bound is also valid for the second product, the expectation of each term in the sum in \eqref{4.21eq} is bounded by~$2\frakq_{L,\beta}(\BB)^{K_j/2}$. The sum over~$(\bt_i)\in\scrC_j(\Y)$ can then be estimated at~$2^{K_j}$ which yields
\begin{equation}
\state{\hatQ_{\theta_{\bt_1}(\GG_i)}\hatQ_{\theta_{\bt_2}(\GG_i^\cc)}}_{L,\beta}
\le\,\, \!\!\!\sum_{\begin{subarray}{c}
\Y\colon \Y\in\scrM_{L/B}\\\bt_1\in\Y,\,\bt_2\not\in\Y
\end{subarray}}
2\,\bigl[4\frakq_{L,\beta}(\BB)\bigr]^{|\Y_j^{\text{ext}}|/2}.
\end{equation}
From here the claim follows by noting that our choice of~$j$ implies~$|\Y_j^{\text{ext}}|\ge\frac1{2d}|\partial\Y|$ (we assume that~$4\frakq_{L,\beta}(\BB)\le1$ without loss of generality).
\end{proofsect}

\begin{proofsect}{Proof of Theorem~\ref{thm-QPT}}
By Lemma~\ref{lemma-qp}, the assumptions on~$\BB$, and \eqref{class-cond} we have that $\frakq_{L,\beta}(\BB)<\delta$. Invoking a standard Peierls argument in toroidal geometry---see, e.g., the proof of \cite[Lemma~3.2]{BCN1}---the right-hand side of \eqref{sum-Lambda} is bounded by a quantity~$\eta(\delta)$ such that~$\eta(\delta)\downarrow0$ as~$\delta\downarrow0$. Choosing~$\delta$ sufficiently small, we will  thus have~$\eta(\delta)\le\epsilon$, proving \eqref{no-clustering}. The bound \eqref{bad-bad} is a consequence of the chessboard estimates which yield~$\langle\hatQ_{\BB}\rangle_{L,\beta}\le\frakq_{L,\beta}(\BB)<\delta$.
\end{proofsect}

\subsection{Exhibiting phase coexistence}
\label{sec4.3}\noindent
In order to complete our general results, we still need to prove Propositions~\ref{prop-SB} and~\ref{prop-EE} whose main point is to guarantee existence of multiple translation-invariant KMS states. (Recall that, throughout this section, we work only with translation-invariant interactions.) Let us refer to
\begin{equation}
\T_L^+=\bigl\{x \in \T_L\, :\, -\floor{\ffrac{L}{4}-\ffrac{1}{2}} \leq x_1\leq \ceil{\ffrac{L}{4}-\ffrac{1}{2}}\bigr\}
\end{equation}
as the ``front side'' of the torus, and to~$\T_L^-$ as the ``back side.'' Let~$\frakA_L^+$ be the~$C^\star$ algebra of all observables localized in~$\T_L^+$ (i.e., an operator in~$\frakA_L^+$ acts as the identity on~$\T_L^-$).

The construction of infinite-volume KMS states will be based on the following standard lemma:

\begin{mylemma}
\label{lemma-KMS}
Let~$\T_{L/B}$ be the factor torus and let~$\Delta_M\subset\T_{L/B}$ be a block of~$M\times\dots\times M$ sites at the ``back side'' of~$\T_{L/B}$ (i.e., we have $\dist(0,\Delta_M)\ge \frac L{2B}-M$). Given a $B$-block event~$\CC$, let
\begin{equation}
\label{hat-rho}
\hat\rho_{L,M}(\CC)=\frac1{|\Delta_M|}\sum_{\bt\in\Delta_M}\hatQ_{\theta_{\bt}(\CC)}.
\end{equation}
Suppose that~$\langle\hatQ_{\CC}\rangle_{L,\beta}\ge c$ for all~$L\gg1$ and some constant~$c>0$, and define the ``conditional'' state~$\state{-}_{L,M;\beta}$ on $\frakA_L^+$ by
\begin{equation}
\label{cond-state}
\state{A}_{L,M;\beta}=\frac{\langle\,\hat\rho_{L,M}(\CC)\, A\rangle_{L,\beta}}{\langle\,\hat\rho_{L,\beta}(\CC)\rangle_{L,\beta}}.
\end{equation}
If~$\state{-}_\beta$ is a (subsequential) weak limit of~$\state{-}_{L,M;\beta}$ as~$L\to\infty$ (along multiples of~$B$) followed by~$M\to\infty$, then~$\state{-}_\beta$ is a KMS state at inverse temperature~$\beta$ which is invariant under translations by~$B$.
\end{mylemma}

\begin{proofsect}{Proof}
Translation invariance is a consequence of ``conditioning'' on the spatially-ave\-raged quantity \eqref{hat-rho}. Thus, all we need to do is to prove that the limit state satisfies the KMS condition \eqref{KMS}. Let~$t\mapsto\alpha_t^{(L)}$ be the unitary evolution on~$\T_L$. If~$B$ is a local observable that depends only on the ``front'' side of the torus the fact that the interaction is finite range and that the series~\eqref{alphat} converges in norm, uniformly in~$L$, implies
\begin{equation}
\label{after-back-of-torus}
\bigl[\alpha_t^{(L)}(B),\hat\rho_{L,M}(\CC)\bigr]\,\underset{L\to\infty}\longrightarrow\,0
\end{equation}
in norm topology, uniformly in~$t$ on compact subsets of~$\C$. 
(Note that, for any $B$ localized inside a fixed finite subset of $\Z^d$, for large enough $L$, it will always be in the ``front'' side $\T_L^+$, under the projection $\Z^d \to \T_L = \Z^d/L\Z^d$.)
This means that for any bounded local operators~$A$ and~$B$ on the ``front'' side of the torus,
\begin{equation}
\state{\,\hat\rho_{L,M}(\CC)\,AB}_{L,\beta}=\state{\,\hat\rho_{L,M}(\CC)\,\alpha_{-\texti\beta}^{(L)}(B)A}_{L,\beta}+o(1),
\qquad L\to\infty.
\end{equation}
(Again, it is no restriction to say that~$A$ and~$B$ are on the ``front'' side, by simply letting $L$ be large enough.)
Since~$\alpha_{-\texti\beta}^{(L)}(B)\to \alpha_{-\texti\beta}(B)$ in norm, the state~$A\mapsto\langle A\rangle_{L,M;\beta}$ converges, as~$L\to\infty$ and~$M\to\infty$, to a KMS state at inverse temperature~$\beta$.
\end{proofsect}

\begin{proofsect}{Proof of Proposition~\ref{prop-SB}}
By $\hatQ_{\BB}+\hatQ_{\GG_1}+\dots+\hatQ_{\GG_n}=\1$, the symmetry assumption and \eqref{bad-bad} we know that
\begin{equation}
\bigl\langle\hatQ_{\GG_k}\bigr\rangle_{L,\beta}\ge\frac{1-\epsilon}n.
\end{equation}
So, if~$\hat\rho_{L,M}(\GG_k)$ is as in \eqref{hat-rho}, the expectation~$\langle\hat\rho_{L,M}(\GG_k)\rangle_{L,\beta}$ is uniformly positive. This means that, for each~$k=1,\dots,n$, we can define the state~$\state{-}^{(k)}_{L,M;\beta}$, $k=1,\dots,n$, by \eqref{cond-state} with the choice~$\CC=\GG_k$. Using \eqref{no-clustering} we conclude
\begin{equation}
\bigl\langle\hatQ_{\theta_{\bt}(\GG_k)}\bigr\rangle^{(k)}_{L,M;\beta}\ge1-\frac{n\epsilon}{1-\epsilon}, \qquad k=1,\dots,n,
\end{equation}
for any~$\bt$ on the ``front'' side of~$\T_{L/B}$ (provided that~$M\ll\ffrac LB$). For~$(n+1)\epsilon<\ffrac12$, the right-hand side exceeds~$\ffrac12$ and so any thermodynamic limit of~$\state{-}^{(k)}_{L,M;\beta}$ as~$L\to\infty$ and~$M\to\infty$ is ``domintated'' by~$\GG_k$-blocks. Since, by Lemma~\ref{lemma-KMS}, any such limit is a KMS state, we have~$n$ distinct states satisfying, as is easy to check, \eqref{3.29}.
\end{proofsect}

\begin{proofsect}{Proof of Proposition~\ref{prop-EE}}
Consider the states~$\state{-}_{L,M;\beta}^{(1)}$ and~$\state{-}_{L,M;\beta}^{(2)}$ defined by \eqref{cond-state} with~$\CC=\GG_1$ and~$\CC=\GG_2$, respectively. From assumption~(1) we know that $a_k\,:=\,\langle\hat\rho_{L,M}(\GG_k)\rangle>0$ for at least one~$k=1,2$ and so, for each~$\beta\in[\beta_1,\beta_2]$, at least one of these states is well defined. We claim that we cannot have~$\langle\hatQ_{\GG_k}^{(k)}\rangle_{L,M;\beta}< 1-4\epsilon$ for both~$k=1,2$. Indeed, if that were the case then
\begin{equation}
\hat\rho_{L,M}(\GG_1)+\hat\rho_{L,M}(\GG_2)+\hat\rho_{L,M}(\BB)=1
\end{equation}
and the bounds \twoeqref{bad-bad}{no-clustering} would yield
\begin{equation}
\begin{aligned}
a_1+a_2&=\bigl\langle\hatQ_{\GG_1}+\hatQ_{\GG_2}\bigr\rangle_{L,\beta}
\\
&=\bigl\langle\hatQ_{\GG_1}\bigr\rangle_{L,M;\beta}^{(1)}\,\bigl\langle\hatQ_{\GG_1}\bigr\rangle_{L,\beta}
+\bigl\langle\hatQ_{\GG_2}\bigr\rangle_{L,M;\beta}^{(2)}\,\bigl\langle\hatQ_{\GG_2}\bigr\rangle_{L,\beta}
\\
&\qquad\quad+\bigl\langle\,\hat\rho_{L,M}(\GG_1)\,\hatQ_{\GG_2}\bigr\rangle_{L,\beta}
+\bigl\langle\,\hat\rho_{L,M}(\GG_2)\,\hatQ_{\GG_1}\bigr\rangle_{L,\beta}
\\
&\qquad\qquad\qquad+\bigl\langle\,\hat\rho_{L,M}(\BB)\,[1-\hatQ_{\BB}]\bigr\rangle_{L,\beta}
\\
&<(1-4\epsilon)(a_1+a_2)+3\epsilon
\end{aligned}
\end{equation}
i.e., $4(a_1+a_2)< 3$. Since $\epsilon\le\ffrac14$ 
this implies~$a_1+a_2<\ffrac34\le1-\epsilon$, in contradiction with assumption~(1).

Hence, we conclude that the larger from $\langle\hatQ_{\GG_k}\rangle^{(k)}_{L,M;\beta}$, $k=1,2$ (among those states that exist) must be at least~$1-4\epsilon$. The same will be true about any thermodynamic limit of these states. Let~$\varXi_k\subset[\beta_1,\beta_2]$, $k=1,2$, be the set of~$\beta\in[\beta_1,\beta_2]$ for which there exists an infinite-volume, translation-invariant KMS state~$\state{-}_\beta$ such that~$\langle\hatQ_{\GG_k}\rangle_\beta\ge1-4\epsilon$. Then~$\varXi_1\cup\varXi_2=[\beta_1,\beta_2]$. Now, any (weak) limit of KMS states for inverse temperatures~$\beta_n\to\beta$ is a KMS state at~$\beta$, and so both~$\varXi_1$ and~$\varXi_2$ are closed. Since~$[\beta_1,\beta_2]$ is closed and connected, to demonstrate a point in~$\varXi_1\cap\varXi_2$ it suffices to show that both~$\varXi_1$ and~$\varXi_2$ are non-empty. For that we will invoke condition~(2) of the proposition: From~$\langle\hatQ_{\GG_1}\rangle_{L,\beta_1}\ge1-2\epsilon$ we deduce
\begin{equation}
\bigl\langle\hatQ_{\GG_1}\bigr\rangle_{L,M;\beta_1}^{(1)}=1-\bigl\langle\hatQ_{\GG_2}+\hatQ_{\BB}\bigr\rangle_{L,M;\beta_1}^{(1)}
\ge1-\frac{2\epsilon}{1-2\epsilon}\ge1-4\epsilon,
\end{equation}
and similarly for~$\langle\hatQ_{\GG_2}\rangle_{L,M;\beta_2}^{(2)}$. Thus~$\beta_1\in\varXi_1$ and~$\beta_2\in\varXi_2$, i.e., both sets are non-empty and so~$\varXi_1\cap\varXi_2\ne\emptyset$ as claimed.
\end{proofsect}

\section{Applications}
\label{sec5}\noindent
Here we will discuss---with varying level of detail---the five quantum models described in the introduction. We begin by listing the various conditions of our main theorems which can be verified without much regard for the particulars of each model. Then, in Sect.~\ref{sec5.2}, we proceed to discuss model~(1) which serves as a prototype system for the application of our technique. Sects.~\ref{sec5.3}-\ref{sec5.5} are devoted to the details specific for models~(2-5). 

\subsection{General considerations}
\label{sec5.1}\noindent
Our strategy is as follows: For each model we will need to apply one of the two propositions from Sect.~\ref{sec3.3}, depending on whether we are dealing with a ``symmetry-breaking'' transition (Proposition~\ref{prop-SB}) or a temperature-driven energy-entropy transition (Proposition~\ref{prop-EE}). The main input we need for this are the inequalities \twoeqref{bad-bad}{no-clustering}. These will, 
in turn, be supplied by Theorem~\ref{thm-QPT}, provided we can check the condition~\eqref{class-cond}. Invoking Theorem~\ref{thm-matrix-element}, which requires that our model satisfies the mild requirements \twoeqref{range}{Lipschitz}, condition~\eqref{class-cond} boils down to showing that~$\frakp_{L,\beta}(\BB)$ is small for the requisite bad event. It is, for the most part, only the latter that needs to be verified on a model-specific basis; the rest can be done in some generality.

\newcommand{\OA}{U_{\text{\rm A}}}
\newcommand{\OB}{U_{\text{\rm B}}}

\smallskip
We begin by checking the most stringent of our conditions: reflection positivity. Here, as alluded to in Remark~\ref{RP-peculiar}, we are facing the problem that reflection positivity may be available only in a particular representation of the model---which is often distinct from that in which the model is \emph{a priori} defined. The ``correct'' representation is achieved by a unitary operation that, in all cases at hand, is a ``product rotation'' of all~spins.

There are two rotations we will need to consider; we will express these by means of unitary operators~$\OA$ and~$\OB$. Consider the Hilbert space~$\HH_{\T_L}=\bigotimes_{\br\in\T_L}[\C^{2\CalS+1}]_{\br}$ and let~$(S_{\br}^x,S_{\br}^y,S_{\br}^z)$ have the usual form---cf~\eqref{S-repre}---on~$\HH_{\T_L}$. In this representation, the action of~$\OA$ on a state $\ket\psi\in\HH_{\T_L}$ is defined by 
\begin{equation}
\OA\ket\psi=\prod_{\br\in\T_L}\texte^{\texti\frac{\pi}{2} S_{\br}^y} \texte^{\texti\frac{\pi}{2} S_{\br}^x} \ket\psi.
\end{equation}
The effect of conjugating by this transformation is the cyclic permutation of the spin components~$S_{\br}^y\to S_{\br}^x\to S_{\br}^z\to S_{\br}^y$. The second unitary, $\OB$, is defined as follows:
\begin{equation}
\OB\ket\psi=\prod_{\begin{subarray}{c}
\br\in\T_L\\\text{odd-parity}
\end{subarray}}
\texte^{\texti\pi S_{\br}^y}
\ket\psi.
\end{equation}
The effect of~$\OB$ on spin operators is as follows: For even-parity~$\br$, the spin operators are as before. For odd-parity~$\br$, the component $S_{\br}^y$ remains the same, while both $S_{\br}^x$ and~$S_{\br}^z$ pick up a minus sign. Here are the precise conditions under which our models are reflection positive (RP):

\begin{mylemma}
\label{lemma5.0}
Let~$\OA$ and~$\OB$ be the unitary transformations defined above. Then:
\begin{enumerate}
\item[(a)]
$\OA H\OA^{-1}$ is RP for models~(4-5), and for model~(2) with $\eusmP(x)=\eusmP_1(x^2)+x\eusmP_2(x^2)$.
\item[(b)]
$\OB H\OB^{-1}$ is RP for models~(1,3).
\item[(c)]
$\OB\OA H\OA^{-1}\OB^{-1}$ is RP for model~(2) with $\eusmP(x)=\eusmP_1(x^2)-x\eusmP_2(x^2)$.
\end{enumerate}
\end{mylemma}

\begin{proofsect}{Proof}
(a) Under the unitary~$\OA$ map, the Hamiltonians of models~(4-5) are only using the $x$ and~$z$-components of the spins, which are both real valued. The resulting interaction couples nearest-neighbor spins ferromagnetically, and thus conforms to~\eqref{RP-Ham}. 

(b) For two-body, nearest-neighbor interactions,~$\OB$ has the effect
\begin{equation}
S_{\br}^\alpha S_{\br'}^\alpha\to-S_{\br}^\alpha S_{\br'}^\alpha,\qquad \alpha=x,z,
\end{equation}
while the $S_{\br}^y S_{\br'}^y$ terms remain unchanged. Writing 
\begin{equation}
S_{\br}^y S_{\br'}^y=-(\texti S^y_{\br})(\texti S^y_{\br'})
\end{equation}
we can thus change the sign of all quadratic terms in the interaction and, at the same time, express all operators by means of real-valued matrices. Under the conditions given in Sect.~\ref{sec1}, the Hamiltonians in \eqref{H1} and \eqref{H3} are then of the desired form~\eqref{RP-Ham}. 

(c) Finally, for model~(2), we first apply the argument in~(a). Then the effect of~$\OB$ is that the minus sign in~$\eusmP(x)=\eusmP_1(x^2)-x\eusmP_2(x^2)$ becomes a plus sign. 
\end{proofsect}

Our next items of general interest are the ``easy'' conditions of Theorem~\ref{thm-matrix-element} and Theorem~\ref{thm-QPT}. These turn out to be quite simple to check:

\begin{mylemma}
\label{lemma5.1}
The transformed versions---as defined in Lemma~\ref{lemma5.0}---of the five models from Sect.~\ref{sec1} satisfy the conditions \twoeqref{range}{Lipschitz} with some finite~$R$ and some~$c_1$ independent of~$\CalS$. Moreover, for each of the models~(1-6) there exists a constant~$C$ such that \eqref{xi-def} holds with~$\xi=C/\CalS$ for all~$\CalS$.
\end{mylemma}

\begin{proofsect}{Proof}
All interactions involve at most two spins so~$R=2$ suffices to have \eqref{range}. Writing the interaction in the form \eqref{Hamiltonian}, the normalization by powers of~$\CalS$ makes the corresponding norms~$\Vert h_\Gamma\Vert$ bounded by a quantity independent of~$\CalS$. This means that \eqref{uniform-bd} holds in any finite set (including the torus, with proper periodic extension of the~$h_\Gamma$'s). As to the Lipschitz bound~\eqref{Lipschitz}, this is the subject of Theorem~2 and Proposition~3 of~\cite{Duffield}. Since~$\CalS^{-1}[\bS_{\br}^\alpha]_\Omega=\Omega_{\br}+O(1/\CalS)$, and similarly for the lower symbol, the same argument proves that~$\xi=O(1/\CalS)$.
\end{proofsect}

To summarize our general observations, in order to apply Propositions~\ref{prop-SB}-\ref{prop-EE}, we only need to check the following three conditions: 
\settowidth{\leftmargini}{(11)}
\begin{enumerate}
\item[(1)]
The requisite bad event is such that~$\vartheta_{\bt}(\BB)=\BB$ for all~$\bt\in\T_{L/B}$.
\item[(2)] 
The occurrence of different types of goodness at neighboring $B$-blocks implies that a block placed in between the two (so that it contains the sites on the boundaries between them)  is bad---cf condition~(2) of Definition~\ref{def-incompatibility}.
\item[(3)]
The quantity~$\frakp_{L,\beta}(\BB)$ is sufficiently small.
\end{enumerate}
In all examples considered in this paper, conditions~(1-2) will be checked directly but condition~(3) will require estimates specific for the model at hand. (Note that, since we are forced to work in the representation that makes the interaction reflection positive; the conditions~(1-3) must be verified in \emph{this} representation.)

\begin{myremark}
It is noted that all of the relevant classical models---regardless of the signs of the interactions---are RP with respect to reflections in planes of sites. We will often use this fact to ``preprocess'' the event underlying~$\frakp_{L,\beta}(\BB)$ by invoking chessboard estimates with respect to these reflections. We will also repeatedly use the subadditivity property of~$\AA\mapsto\frakp_{L,\beta}(\AA)$ as stated in~\cite[Theorem~6.3]{BCN1}. Both of these facts will be used without (much) apology.
\end{myremark}


\subsection{Anisotropic Heisenberg antiferromagnet}
\label{sec5.2}\noindent
Consider the reflection-positive version of the Hamiltonian~\eqref{H1} which (in the standard representation of the spin operators) on the torus~$\T_L$ takes the form
\begin{equation}
H_L=-\sum_{\langle\br,\br'\rangle}\CalS^{-2}
(J_1 S^x_{\br}S^x_{\br'}-J_2S^y_{\br}S^y_{\br'}+S^z_{\br}S^z_{\br'}).
\end{equation}
(The classical version of~$H_L$ is obtained by replacing each~$S_{\br}^{\alpha}$ by the corresponding component of~$\CalS\Omega_{\br}$.)
The good block events will be defined on a $2\times\dots\times2$ block~$\Lambda_B$---i.e.,~$B=2$---and, roughly speaking, they will represent the two \emph{ferromagnetic} states in the~$z$-direction one can put on~$\Lambda_B$. Explicitly, let~$\GG_+$ be the event that~$\Omega_{\br}=(\theta_{\br},\phi_{\br})$ satisfies~$|\theta_{\br}|<\kappa$ for al~$\br\in\Lambda_B$ and let~$\GG_-$ be the event that~$|\theta_{\br}-\pi|<\kappa$ for all~$\br\in\Lambda_B$.

\begin{theorem}[Heisenberg antiferromagnet]
\label{thm-AF}
Let~$d\ge2$ and let~$0\le J_1,J_2<1$ be fixed. For each~$\epsilon>0$ and each~$\kappa>0$, there exist constants~$c$ and~$\beta_0$ and, for all~$\beta$ and~$\CalS$ with~$\beta_0\le\beta\le c\sqrt\CalS$, there exist two distinct, translation-invariant KMS states~$\state{-}^+_\beta$ and~$\state{-}_\beta^-$ with the property
\begin{equation}
\bigl\langle\hatQ_{\GG_\pm}\bigr\rangle_\beta^\pm\ge1-\epsilon.
\end{equation}
In particular, for all such~$\beta$ we have
\begin{equation}
\state{S_0^z}_\beta^+-\state{S_0^z}_\beta^->0.
\end{equation}
\end{theorem}

\begin{proofsect}{Proof}
Let~$\BB=(\GG_+\cup\GG_-)^\cc$ be the bad event. It is easy to check that~$\vartheta_{\bt}$ acts on~$\BB$ only via translations. Moreover, if~$\GG_+$ and~$\GG_-$ occur at neighboring (but disjoint) translates of~$\Lambda_B$, then the block between these is necessarily bad. In light of our general observations from Sect.~\ref{sec5.1}, we thus only need to produce good bounds on~$\frakp_{L,\beta}(\BB)$, the classical probability of bad behavior. Since these arguments are standard and appear, for all intents and purposes, in the union of Refs.~\cite{FILS1,FILS2,Senya,CKS}, we will be succinct (and not particularly efficient).

Let $\Delta=\min\{(1-J_1),(1-J_2),2/a_d\}$ where~$a_d=d2^{d-1}$ and fix~$\eta>0$ with~$\eta\ll1$ such that
\begin{equation}
\label{eta-cond}
1-\cos\eta-\Delta\sin^2\kappa<0.
\end{equation}
We will start with a lower estimate on the full partition function. For that we will restrict attention to configurations where~$|\theta_{\br}|\le\ffrac\eta2$ for all~$\br\in\T_L$. The interaction energy of a pair of spins is clearly maximized when both the~$x$ and~$y$-terms are negative. This allows us to bound the energy by that in the isotropic case~$J_1=J_2=1$---i.e., the cosine of the angle between the spins. Hence, the energy between each neighboring pair is at most~$(-\cos\eta)$. We arrive at
\begin{equation}
Z_L(\beta)\ge \bigl[V(\eta)\texte^{d\beta\cos\eta}\bigr]^{L^d},
\end{equation}
where the phase volume~$V(\eta)=2\pi[1-\cos(\ffrac\eta2)]$ may be small but is anyway independent of~$\beta$.

To estimate the constrained partition function in the numerator of~$\frakp_{L,\beta}(\BB)$, we will classify the bad blocks into two distinct categories: First there will be blocks where not all spins are within~$\kappa$ of the pole and, second, there will be those bad blocks which, notwithstanding their Ising nature, will have defects in their ferromagnetic pattern. We denote the respective events by~$\BB_1$ and~$\BB_2$. To bound~$\frakp_{L,\beta}(\BB_1)$, since we may decorate the torus from a single site, we may as well run a single site argument~$2^d$-times. We are led to consider the constrained partition function where every site is outside its respective polar cap. It is not hard to see that the maximal possible interaction is $1-\Delta\sin^2\kappa$; we may estimate the measure of such configurations as full. Thus,
\begin{equation}
\frakp_{L,\beta}(\BB_1)\le 2^d\,\frac{4\pi}{V(\eta)}\texte^{\,\beta d(1-\cos\eta-\Delta\sin^2\kappa)}.
\end{equation}
Note that, by \eqref{eta-cond}, this is small when~$\beta\gg1$.

The less interesting Ising violations are estimated as follows: The presence of such violations implies the existence of a bond with nearly antialigned spins. We estimate the interaction of this bond at~$\cos(2\kappa)$. Now there are~$a_d$ bonds on any cube so when we disseminate---using reflections through sites---we end up with at least one out of every~$a_d$ bonds with this energy. The rest we may as well assume are fully ``aligned''---and have energy at least negative one---and we might as well throw in full measure, for good measure. We thus arrive at
\begin{equation}
\frakp_{L,\beta}(\BB_2)\le a_d\,
\frac{4\pi}{V(\eta)}\exp\biggl\{\beta d\Bigl(\frac1{a_d}\cos(2\kappa)+1-\frac1{a_d}-\cos\eta\Bigr)\biggr\}
\end{equation}
as our estimate for each such contribution to the Ising badness. Here the prefactor~$a_d$ accounts for the choice of the ``bad'' bond. Since~$1/a_d>\ffrac\Delta2$, the constant multiplying~$\beta d$ in the exponent is less than the left-hand side of \eqref{eta-cond}; hence~$\frakp_{L,\beta}(\BB_2)\ll1$ once~$\beta\gg1$ as well. It follows that, given~$J_1,J_2<1$, we can find~$\beta_0$ sufficiently large so that
$\frakp_{L,\beta}(\BB)\le\frakp_{L,\beta}(\BB_1)+\frakp_{L,\beta}(\BB_2)\ll1$ once~$\beta\ge\beta_0$. The statement of the theorem is now implied by Proposition~\ref{prop-SB} and the~$\pm$-symmetry of the model.
\end{proofsect}


\subsection{Large-entropy models}
\label{sec5.3}\noindent
Here we will state and prove order-disorder transitions in models~(2-3).  As in the previous subsection, most of our analysis is classical.  While we note that much of the material of this section has appeared in some form before, e.g., in \cite{Dobrushin-Shlosman,Kotecky-Shlosman,Senya,CKS,Senya_VE-I,vES}, here we must go a slightly harder route dictated by the quantum versions of reflection positivity. 

We start with the observation that model~(2) with~$\eusmP(x)=\eusmP_1(x^2)-x\eusmP_2(x^2)$ is unitarily equivalent, via a rotation of all spins about the~$z$-axis, to the same model with~$\eusmP(x)=\eusmP_1(x^2)+x\eusmP_2(x^2)$. Hence, it suffices to consider only the case of the plus sign. We thus focus our attention on models with classical Hamiltonians of the~form
\begin{equation}
\label{LE-Ham}
H^\infty(\Omega)=
-\sum_{\langle {\br},{\br'} \rangle}
\sum_{k=1}^p c_k\,(\Omega_{\br} \diamond \Omega_{\br'})^k,
\qquad c_k\ge0,
\end{equation}
where $(\Omega_1\diamond \Omega_2)$ denotes the variant of the usual dot product~$\Omega_1^{(x)}\Omega_1^{(x)}-\Omega_1^{(y)}\Omega_1^{(y)}+\Omega_1^{(z)}\Omega_1^{(z)}$ for model~(3), and the ``dot product among the first two components'' for model~(2). We now state our assumptions which ensure that models (2) and (3) have the large entropy property.

Let us regard the coefficients in \eqref{LE-Ham} as an infinite (but summable) sequence, generally thought of as terminating when $k=p$. (For the most part we \emph{will} require that~$\frakE_p$ be a polynomial. However, some of our classical calculations apply even for genuine power series.) The terms of this sequence may depend on~$p$ so we will write them as $\frakc^{(p)}=(c_1^{(p)},c_2^{(p)}, \dots)$; we assume that the $\ell^1$-norm of each~$\frakc^{(p)}$ is one. Let $\frakE_p\colon[-1,1]\to\R$ be defined by
\begin{equation}
\frakE_p(x) = \sum_{k\ge1} c_k^{(p)}x^k.
\end{equation}
Here is the precise form of the large-entropy property:

\begin{mydefinition}
\label{def-LEP}
We say that the sequence~$(\frakc^{(p)})$ has the \emph{large entropy property} if there is a sequence~$(\epsilon_p)$ of positive numbers with $\epsilon_p \downarrow 0$ such that the functions 
\begin{equation}
A_p(s)  =  \frakE_p(1-\epsilon_p s)
\end{equation}
converge---uniformly on compact subsets of $[0,\infty)$---to a function $s\mapsto A(s)$ with 
\begin{equation}
\lim_{s\to 0^+} A(s) = 1
\quad\text{and}\quad
\lim_{s \to \infty} A(s) = 0
\end{equation}
\end{mydefinition}

\begin{myremark}
Despite the abstract formulation, the above framework amalgamates all known examples~\cite{Senya_VE-I,vES} and provides plenty of additional generality. A prototypical example that satisfies Definition~\ref{def-LEP} is the sequence arising as the coefficients of the polynomial $\frakE_p(x)=(\frac{1+x}2)^p$. A general class of sequences ${\frakc}^{(p)}$ is defined from a probability density function $\phi\colon[0,1]\to[0,\infty)$ via~$c_k^{(p)}=\frac1p\phi(\ffrac kp)$. In these cases we can generically take $\epsilon_p = \ffrac1p$ and the limiting function~$A$ is then given by~$A(s)=\int_0^1\phi(\lambda)\texte^{-\lambda s}\textd\lambda$. However, as the example $\frakE_p(x)=(\frac{1+x}2)^p$ shows, existence of such a density function is definitely not a requirement for the large-entropy property to hold. What is required is that the ``distribution function'' $\sum_{k\le ps}c_k^{(p)}$ is small for~$s\ll1$.
\end{myremark}

Our analysis begins with the definition of good and bad events. 
First we will discuss the situation on bonds: The bond $\langle\br,\br'\rangle$ is considered to be \emph{energetically 
good} if the attractive energy is larger (in magnitude) than some strictly positive constant~$b$ (a number of order unity depending on gross details, where we recall that 1 is the optimal value), i.e., if
\begin{equation}
\frakE_p(\Omega_{\br}\diamond \Omega_{\br'}) \geq b.
\end{equation}
The \emph{entropically good} bonds are simply the complementary events (so that every bond is a good bond).  Crucial to the analysis is the fact, ensured by our large entropy assumption, that the crossover between the energetic and entropic phenotypes occurs when the deviation between neighboring spins is of the order $\sqrt{\epsilon_p}$.

We define the good block events~$\Gso$ and~$\Gsd$ on the $2\times\dots\times2$-block~$\Lambda_B$ as follows: $\Gso$ is the set of spin configurations where every bond on~$\Lambda_B$ is energetically good while~$\Gsd$ collects all spin configurations where every bond on~$\Lambda_B$ is entropically good. The requisite bad event is defined as~$\BB=(\Gso\cup\Gsd)^\cc$.

\smallskip
Our fundamental result will be a proof that the density of energetically good blocks is discontinuous:

\begin{theorem}[Large-entropy models]
\label{thm-LE}
Consider a family of finite sequences $\frakc^{(p)}=(c_k^{(p)})_{k\le p}$ and suppose that~$\frakE_p$ have the large entropy property in the sense of Definition~\ref{def-LEP}. Consider the quantum spin systems with the Hamiltonian 
\begin{equation}
\label{Hp}
H^{(p)}=-\sum_{\langle\br,\br'\rangle}\frakE_p\bigl(\CalS^{-2}(S_{\br}\diamond S_{\br'})\bigr),
\end{equation}
(with both interpretations of $(S_{\br}\diamond S_{\br'})$ possible). Then there exists $b\in(0,1)$ for which the associated energetic bonds have discontinuous density in the large $\CalS$ quantum systems. Specifically, for every $\epsilon>0$ there is a $p_0<\infty$ so that for any $p>p_0$ and all~$\CalS$ sufficiently large, there is an inverse temperature~$\betat$ at which there exist two distinct, translation-invariant KMS states $\state{-}_{\betat}^\ord$ and $\state{-}_{\betat}^\dis$ with the property
\begin{equation}
\label{LE-transition}
\langle\hatQ_{\Gso}\rangle_{\betat}^\ord\ge1-\epsilon
\quad\text{and}\quad
\langle\hatQ_{\Gsd}\rangle_{\betat}^\dis\ge1-\epsilon.
\end{equation}
\end{theorem}

With a few small additional ingredients, we show that the above implies that the energy density itself is discontinuous:

\begin{mycorollary}
\label{cor-Discont}
There exist constants $b$ and $b'$, both strictly less than $\ffrac12$, such that 
the energy density~$\frake(\beta)$---defined via the~$\beta$-derivative of the free energy---satisfies
\begin{equation}
\label{e-beta-jump}
\frake(\beta)\begin{cases}
\ge1-b',\qquad&\text{if }\beta>\betat,
\\*[1mm]
\le b,\qquad&\text{if }\beta<\betat,
\end{cases}
\end{equation}
for all $p$ sufficiently large.
\end{mycorollary}

The bulk of the proof of this theorem again boils down to the estimate of~$\frakp_{L,\beta}(\BB)$:

\begin{myproposition}
\label{prop-LE}
There exist~$b_0\in(0,1)$, $\Delta>0$, $C<\infty$, and for each $b \in (0,b_0]$ there exists~$p_0<\infty$ such that
\begin{equation}
\label{5.21two}
\lim_{L\to\infty}\frakp_{L,\beta}(\BB)<C(\epsilon_p)^{\Delta}
\end{equation}
hold for all~$p\ge p_0$ and all~$\beta\ge0$.
\end{myproposition}

Apart from a bound on~$\frakp_{L,\beta}(\BB)$, we will also need to provide the estimates in condition~(2) of Proposition~\ref{prop-EE}. Again we state these in their classical form:

\begin{myproposition}
\label{prop5.10}
There exist constants~$C_1<\infty$, $p_1<\infty$ and $\Delta_1>0$ such that the following is true for all~$p\ge p_1$: First, at~$\beta=0$ we have
\begin{equation}
\label{zero-bd}
\limsup_{L\to\infty}\frakp_{L,0}(\Gso)\le C_1 (\epsilon_p)^{\Delta_1}.
\end{equation}
Second, if~$\beta_0\in(0,\infty)$ is large enough, specifically if $\texte^{\beta_0 d} \geq \epsilon_p^{-2(1+\Delta_1)}$, then
\begin{equation}
\label{beta-zero-bd}
\limsup_{L\to\infty}\frakp_{L,\beta_0}(\Gsd)\le C_1 (\epsilon_p)^{\Delta_1}.
\end{equation}
\end{myproposition}

The proof of these propositions is somewhat technical; we refer the details to the Appendix, where we will also prove the corollary.

\begin{proofsect}{Proof of Theorem~\ref{thm-LE}}
We begin by verifying the three properties listed at the end of Sect.~\ref{sec5.1}.
As is immediate from the definitions, neighboring blocks of distinct type of goodness must be separated by a bad block. Similarly, reflections~$\theta_{\bt}$ act on~$\BB$ only as translations. To see that the same applies to the ``complex'' reflections~$\vartheta_{\bt}$, we have to check that~$\BB$ is invariant under the ``complex conjugation'' map~$\sigma$. For that it suffices to verify that~$\sigma(\Omega)\diamond\sigma(\Omega')=\Omega\diamond\Omega'$ for any~$\Omega,\Omega'\in\scrS_2$. This follows because both interpretations of~$\Omega\diamond\Omega$ are quadratic in the components of~$\Omega$ and because~$\sigma$ changes the sign of the~$y$-component and leaves the other components intact.

Let~$b<b_0$\, where~$b_0$ is as in Proposition~\ref{prop-LE}. 
Then \eqref{5.21two} implies that~$\frakp_{L,\beta}(\BB)\ll1$ once $\epsilon_p\ll1$.  Quantum chessboard estimates yield~$\langle\hatQ_{\AA}\rangle_{L,\beta}\le\frakq_{L,\beta}(\AA)$ which by means of Theorem~\ref{thm-matrix-element} implies that both~$\langle\hatQ_{\Gsd}\rangle_{L,0}$ and~$\langle\hatQ_{\Gso}\rangle_{L,\beta_0}$ are close to one once~$L\gg1$ and~$\sqrt{\CalS}$ is sufficiently large compared with $\beta_0$ (referring to Proposition~\ref{prop5.10}). Theorem~\ref{thm-QPT} then provides the remaining conditions required for application of Proposition~\ref{prop-EE}; we conclude that there exists a~$\betat\in[0,\beta_0]$ and two translation-invariant KMS states $\state{-}_{\betat}^\ord$ and $\state{-}_{\betat}^\dis$ such that \eqref{LE-transition} hold.
\end{proofsect}

\begin{myremarks}
Again, a few remarks are in order:
\settowidth{\leftmargini}{(11)}
\begin{enumerate}
\item[(1)]
Note that the theorem may require larger~$\CalS$ for larger~$p$, even though in many cases the transition will occur uniformly in~$\CalS\gg1$ once~$p$ is sufficiently large. The transition temperature~$\betat$ will generally depend on~$p$ and~$\CalS$.
\item[(2)]
There are several reasons why Theorem~\ref{thm-LE} has been stated only for polynomial interactions. First, while the upper symbol is easily---and, more or less, unambi\-guously---defined for polynomials, its definition for general functions may require some non-trivial limiting procedures that have not been addressed in the literature. Second, the reduction to the classical model, cf Corollary~\ref{cor3.2}, requires that the classical interaction be Lipschitz, which is automatic for polynomials but less so for general power series. In particular, Theorem~\ref{thm-LE} does not strictly apply to non-smooth (or even discontinuous) potentials even though we believe that, with some model-specific modifications of the proof of Theorem~\ref{thm-matrix-element}, we could include many such cases as well.
\end{enumerate}

\end{myremarks}


\subsection{Order-by-disorder transitions: Orbital-compass model}
\label{sec5.4}\noindent
We begin by the easier of the models~(4-5), the 2D orbital compass model. We stick with the reflection-positive version of the Hamiltonian which, on~$\T_L$, is given by
\begin{equation}
\label{OCM}
H_L=-\CalS^{-2}\sum_{\br\in\T_L}\sum_{\alpha=x,z}S_{\br}^{(\alpha)}S_{\br+\hate_\alpha}^{(\alpha)},
\end{equation}
with~$\hate_x,\hate_y,\hate_z$ denoting the unit vectors in (positive) coordinate directions.
The number~$B$ will only be determined later, so we define the good events for general~$B$. Given~$\kappa>0$ (with~$\kappa\ll1$), let~$\GG_x$ be the event that all (classical) spins on a $B\times B$ block~$\Lambda_B$ satisfy
\begin{equation}
|\Omega_{\br}\cdot\hate_x|\ge\cos(\kappa).
\end{equation}
Let~$\GG_z$ be the corresponding event in the~$z$ spin-direction. Then we have:

\begin{theorem}[Orbital-compass model]
\label{thm-OCM}
Consider the model with the Hamiltonian as in~\eqref{H4}. For each~$\epsilon>0$ there exist~$\kappa>0$,~$\beta_0>0$ and~$c>0$ and, for each~$\beta$ with~$\beta_0\le\beta\le c\sqrt\CalS$, there is a positive integer~$B$ and two distinct, translation-invariant KMS states~$\state{-}_\beta^{(x)}$ and~$\state{-}_\beta^{(z)}$ such that
\begin{equation}
\label{OCM-QQ}
\bigl\langle\hatQ_{\GG_\alpha}\bigr\rangle^{(\alpha)}_\beta\ge1-\epsilon,\qquad \alpha=x,z.
\end{equation}
In particular, for all~$\beta$ with~$\beta_0\le\beta\le c\sqrt\CalS$,
\begin{equation}
\label{OCM-SS}
\bigl\langle (\bS_{\br}\cdot\hate_\alpha)^2\bigr\rangle^{(\alpha)}_\beta\ge \CalS^2(1-\epsilon),
\qquad \alpha=x,z.
\end{equation}
\end{theorem}

The proof is an adaptation of the results from \cite{BCN1,BCN2,BCKiv} for the classical versions of order-by-disorder. Let~$\BB=(\GG_x\cup\GG_z)^\cc$ denote the requisite bad event. By definition,~$\BB$ is invariant under reflections of (classical) spins through the $xz$-plane; i.e.,~$\sigma(\BB)=\BB$. Since the restrictions from~$\BB$ are uniform over the sites in~$\Lambda_B$, we have~$\vartheta_{\bt}(\BB)=\tau_{B\bt}(\BB)$. So, in light of our general claims from Sect.~\ref{sec5.1}, to apply the machinery leading to Proposition~\ref{prop-SB}, it remains to show that~$\frakp_{L,\beta}(\BB)$ is small if~$\beta\gg1$ and the scale~$B$ is chosen appropriately. For that let~$H^\infty(\Omega)$ denote the classical version of the Hamiltonian \eqref{OCM}. By completing the nearest-neighbor terms to a square, we get
\begin{equation}
\label{OCMclass}
H^\infty(\Omega)=\frac12\sum_{\br\in\T_L}\sum_{\alpha=x,z}(\Omega_{\br}^{(\alpha)}-\Omega_{\br+\hate_\alpha}^{(\alpha)})^2+\sum_{\br\in\T_L}[\Omega_{\br}^{(y)}]^2-|\T_L|.
\end{equation}
Here~$\Omega_{\br}^{(\alpha)}$ denotes the~$\alpha$-th Cartesian component of~$\Omega_{\br}$.

Unforuntately, the event~$\BB$ is too complex to allow a direct estimate of~$\frakp_{L,\beta}(\BB)$. Thus, we will decompose~$\BB$ into two events,~$\BBE$ and~$\BBSW$ depending on whether the ``badness'' comes from bad energy or bad entropy. Let~$\Delta>0$ be a scale whose size will be determined later. Explicitly, the event~$\BBE$ marks the situations that either
\begin{equation}
|\Omega_{\br}^{(y)}|\ge c_1\Delta
\end{equation}
for some site~$\br\in\Lambda_B$, or
\begin{equation}
\label{E:5.25}
|\Omega_{\br}^{(a)}-\Omega_{\br+\hate_{\alpha}}^{(\alpha)}|\ge c_2\Delta/B,
\end{equation}
for some pair $\br$ and $\br+\hate_{\alpha}$, both in $\Lambda_B$.
Here~$c_1,c_2$ are constants to be determined momentarily.
The event~$\BBSW$ is simply given by
\begin{equation}
\BBSW=\BB\setminus\BBE.
\end{equation}
By the subadditity property of~$\frakp_{L,\beta}$, we have~$\frakp_{L,\beta}(\BB)\le\frakp_{L,\beta}(\BBE)+\frakp_{L,\beta}(\BBSW)$.

Since~$\BBE$ implies the existence of an energetically ``charged'' site or bond with energy about~$(\ffrac\Delta B)^2$ above its minimum, the value of~$\frakp_{L,\beta}(\BBE)$ is estimated relatively easily:
\begin{equation}
\label{pBBE}
\frakp_{L,\beta}(\BBE)\le c\beta B^2\texte^{-\tilde c\beta\Delta^2/B^2},
\end{equation}
for some constants~$c$ and~$\tilde c$. (Here~$cB^2$ accounts for possible positions of the ``excited'' bond/site and~$\beta$ comes from the lower bound on the classical partition function.) 

As to~$\BBSW$, here we will decompose further into more elementary events: Given a collection of 
vectors~$\hatw_1,\dots,\hatw_s$ that are uniformly spaced on the first quadrant of the main circle, $\scrS_1^{++}=\{\Omega\in\scrS_2\colon\Omega\cdot\hate_y=0,\Omega^{(x)}\geq 0, \Omega^{(z)}\geq 0\}$, we define~$\BBSW^{(i)}$ to be the set of configurations in~$\BBSW$ such that
\begin{equation}
|\Omega_{\br}^{(x)}\cdot\hatw_i^{(x)}| + |\Omega_{\br}^{(z)}\cdot\hatw_i^{(z)}| 
\ge\cos(\Delta),
\qquad\br\in\Lambda_B.
\end{equation}
Since~$\BBSW$ is disjoint from~$\BBE$, on~$\BBSW$ the $y$-component of every spin is less than order~$\Delta$ and any neighboring pair of spins differ by angle at most~$\Delta$ (up to a reflection). Hence, by choosing~$c_1$ and~$c_2$ appropriately, any two spins in~$\Lambda_B$ will differ by less than~$\Delta$ from some~$\hatw_i$, i.e.,
\begin{equation}
\BBSW\subset\bigcup_{i=1}^s\BBSW^{(i)},
\end{equation}
provided that $s \Delta$ exceeds  the total length of~$\scrS_1^{++}$. To estimate~$\frakp_{L,\beta}(\BBSW^{(i)})$ we will have to calculate the constrained partition function for the event~$\BBSW^{(i)}$. The crucial steps of this estimate are encapsulated into the following three propositions:

\begin{myproposition}
\label{prop-SWbd}
Consider the classical orbital compass model with the Hamiltonian $H^\infty(\Omega)$ as in \eqref{OCMclass} and suppose that~$\Delta\ll1$. Then for all~$i=1,\dots,s$,
\begin{equation}
\label{SWbd}
\frakp_{L,\beta}(\BBSW^{(i)})\le 2^{2B}\texte^{-B^2(F_{L,\Delta}(\hatw_i)-F_{L,\Delta}(\hate_1))},
\end{equation}
where, for each~$\hatw\in\scrS_1^{++}=\{\hatv\in\scrS_2\colon\hatv\cdot\hate_2=0,\hatv^{(x)}\ge 0,\hatv^{(z)}\ge 0\}$,
\begin{equation}
\label{FLDelta}
F_{L,\Delta}(\hatw)=-\frac1{L^2}\log\int_{(\scrS_2)^{|\T_L|}}\!\!\textd\Omega\,\,\Bigl(\frac{\beta\texte^{\beta}}{2\pi}\Bigr)^{|\T_L|}\,\texte^{-\beta H^\infty(\Omega)}\,\biggl(\,\prod_{\br\in\T_L}\1_{\{\Omega_{\br}\cdot\hatw\ge\cos(\Delta)\}}\biggr).
\end{equation}
\end{myproposition}

\begin{myproposition}
\label{prop-FElim} 
For each~$\epsilon>0$ there exists~$\delta>0$ such that if
\begin{equation}
\label{beta-Delta}
\beta\Delta^2>\frac1\delta\quad\text{\rm and}\quad\beta\Delta^3<\delta,
\end{equation}
then for all~$L$ sufficiently large, $|F_{L,\Delta}(\hatw)-F(\hatw)|<\epsilon$ holds for any~$\hatw\in\scrS_1^{++}$ with~$F$ given by
\begin{equation}
F(\hatw)=\frac12\int_{[-\pi,\pi]^2}\frac{\textd\bk}{(2\pi)^2}\log\wh D_{\bk}(\hatw).
\end{equation}
Here~$\wh D_{\bk}(\hatw)=\hatw_z^2|1-\texte^{\texti k_1}|^2+\hatw_x^2|1-\texte^{\texti k_2}|^2$.
\end{myproposition}

\begin{myproposition}
\label{prop-FEmin}
The function~$\hatw\mapsto F(\hatw)$ is minimized (only) by vectors~$\hatw=\pm\hate_x$ and~$\hatw=\pm\hate_z$.
\end{myproposition}

The proofs of these propositions consist of technical steps which are deferred to the Appendix.
We now finish the formal proof of the theorem subject to these propositions:

\begin{proofsect}{Proof of Theorem~\ref{thm-OCM} completed}
As already mentioned, the bad event is invariant under both spatial reflections~$\theta_{\bt}$ and the ``internal'' reflection~$\sigma$; hence~$\vartheta_{\bt}(\BB)=\tau_{B\bt}(\BB)$ as desired.
Second, if two distinct good events occur in neighboring blocks, say~$\Lambda_B$ and~$\Lambda_B+B\hate_1$, then at least one of the bonds between these blocks must obey \eqref{E:5.25}; i.e., the box~$\Lambda_B+\hate_1$ is (energetically) bad. Third, we need to show that~$\frakp_{L,\beta}(\BB)$ is small. We will set~$\Delta$ and~$B$ to the values
\begin{equation}
\label{B-Delta}
\Delta=\beta^{-\frac5{12}}
\quad\text{and}\quad
B\approx\log\beta.
\end{equation}
These choices make~$\frakp_{L,\beta}(\BBE)$ small once~$\beta$ is sufficiently large and, at the same time, ensure that \eqref{beta-Delta} holds for any given~$\delta$. Since we have \eqref{SWbd},
Propositions~\ref{prop-FElim}-\ref{prop-FEmin} and the fact that~$\BBSW^{(i)}$, being a subset of~$\BB$, is empty when~$\hatw_i$ is within, say,~$\ffrac\kappa2$ of~$\pm\hate_x$ or~$\pm\hate_z$ tell us that
\begin{equation}
\frakp_{L,\beta}(\BBSW)\le s\texte^{-\frac12\epsilon B^2}
\end{equation}
once~$B$ is sufficiently large. But~$s$ is proportional to $\ffrac1\Delta$ and so this is small for~$\beta$ sufficiently large. We conclude that as~$\beta\to\infty$, we have $\frakp_{L,\beta}(\BB)\to0$ for the above choice of~$B$ and~$\Delta$.

Having verified all required conditions, the $xz$-symmetry of the model puts us in a position to apply Proposition~\ref{prop-SB}. Hence, for all sufficiently large~$\beta$, there exist two infinite-volume, translation-invariant KMS states~$\state{-}_\beta^{(x)}$ and $\state{-}_\beta^{(z)}$ such that \eqref{OCM-QQ} holds. To derive \eqref{OCM-SS}, we note that, for any vector~$\hatw\in\scrS_2$ and any single-spin coherent state~$\ket\Omega$
\begin{equation}
\bS\cdot\hatw\ket\Omega=\CalS(\hatw\cdot\Omega)\ket\Omega+O(\sqrt\CalS).
\end{equation}
Hence, $(\bS\cdot\hate_k)^2\hatQ_{\GG_k}=\CalS^2\hatQ_{\GG_k}+O(\CalS^{3/2})$, where all error terms indicate bounds in norm. Invoking \eqref{OCM-QQ}, the bound \eqref{OCM-SS} follows.
\end{proofsect}

\begin{myremark}
The 3D orbital-compass model is expected to undergo a similar kind of symmetry breaking, with three distinct states ``aligned'' along one of the three lattice directions. However, the actual proof---for the classical model, a version of this statement has been established in~\cite{BCN2}---is considerably more involved because of the existence of (a large number of) inhomogeneous ground states that are not distinguished at the leading order of spin-wave free-energy calculations.
We also note that an independent analysis of the classical version of the 2D orbital-compass model, using an approach similar to Refs.~\cite{BCN1,BCN2} and~\cite{EPL-kompasy}, has been performed in~\cite{Mishra-banda}.
\end{myremark}


\subsection{Order-by-disorder transitions: 120-degree model}
\label{sec5.5}\noindent
The statements (and proofs) for the 120-degree model are analogous, though more notationally involved. Consider six vectors~$\hatv_1,\dots,\hatv_6$ defined by
\begin{alignat}{5}
\hatv_1&=\hate_x,&\qquad\hatv_2&=\tfrac12\hate_x+\tfrac{\sqrt3}2\hate_z,&\qquad
\hatv_3&=-\tfrac12\hate_x-\tfrac{\sqrt3}2\hate_z
\\
\hatv_4&=-\hate_x,&\qquad\hatv_5&=-\tfrac12\hate_x-\tfrac{\sqrt3}2\hate_z,&\qquad
\hatv_6&=\tfrac12\hate_x-\tfrac{\sqrt3}2\hate_z.
\end{alignat}
As is easy to check, these are the six sixth complex roots of unity.
The reflection-positive version of the Hamiltonian on~$\T_L$ then has the form
\begin{equation}
\label{120M}
H=-\CalS^{-2}\sum_{\br\in\T_L}\sum_{\alpha=1,2,3}(\bS_{\br}\cdot\hatv_{2\alpha})(\bS_{\br+\hate_\alpha}\cdot\hatv_{2\alpha}),
\end{equation}
where~$\hate_1,\hate_2,\hate_3$ is yet another labeling of the usual triplet of coordinate vectors in~$\Z^3$.
To define good block events, let~$\kappa>0$ satisfy~$\kappa\ll1$ and let~$\GG_1,\dots,\GG_6$ be the $B$-block events that all spins~$\Omega_{\br}$, $\br\in\Lambda_B$, are such that
\begin{equation}
\Omega_{\br}\cdot\hatv_\alpha\ge\cos(\kappa),
\qquad\alpha=1,\dots,6,
\end{equation}
respectively. Then we have:

\begin{theorem}[120-degree model]
\label{thm-120}
Consider the 120-degree model with the Hamiltonian~\eqref{120M}. For each~$\epsilon>0$ there exist~$\kappa>0$, $\beta_0>0$ and~$c>0$ and, for each~$\beta$ with~$\beta_0\le\beta\le c\sqrt\CalS$, there is a number~$B$ and six distinct, translation-invariant states~$\state{-}_\beta^{(\alpha)}$, $\alpha=1,\dots,6$, such that
\begin{equation}
\label{QQ-120}
\bigl\langle\hatQ_{\GG_\alpha}\bigr\rangle^{(\alpha)}_\beta\ge1-\epsilon,\qquad \alpha=1,\dots,6.
\end{equation}
In particular, for all~$\beta$ with~$\beta_0\le\beta\le c\sqrt\CalS$,
\begin{equation}
\label{spin-120}
\bigl\langle\, \bS_{\br}\cdot\hatv_\alpha\bigr\rangle^{(\alpha)}_\beta\ge \CalS(1-\epsilon),
\qquad \alpha=1,\dots,6.
\end{equation}
\end{theorem}

Fix~$\kappa>0$ (with~$\kappa\ll1$) and let~$B$ and~$\Delta$ be as in \eqref{B-Delta}. Let~$\BB=(\GG_1\cup\dots\cup\GG_{6})^{\cc}$ be the relevant bad event. It is easy to check that~$\BB$ is invariant with respect to~$\sigma$ and, consequently,~$\vartheta_{\bt}(\BB)=\BB$ for all~$\br\in\T_{L/B}$ as required. Introducing the projections
\begin{equation}
\Omega_{\br}^{(\alpha)}=\Omega_{\br}\cdot\hatv_{\alpha},
\qquad\alpha=1,\dots,6,
\end{equation}
and noting that, for any vector~$\hatw\in\scrS_2$,
\begin{equation}
\sum_{\alpha=1,2,3}(\hatw\cdot\hatv_{\alpha})^2=\frac32\bigl[1 -(\hatw\cdot\hate_y)^2\bigr],
\end{equation}
the classical Hamiltonian~$H^\infty(\Omega)$ can be written in the form
\begin{equation}
\label{120-class}
H^\infty(\Omega)=
\frac12\sum_{\br\in\T_L}\sum_{\alpha=1,2,3}(\Omega_{\br}^{(2\alpha)}-\Omega_{\br+\hate_{\alpha}}^{(2\alpha)})^2
+\frac32\sum_{\br\in\T_L}(\Omega_{\br}\cdot\hate_y)^2-\frac{3}{2}|\T_L|.
\end{equation}
As for the orbital-compass model, we will estimate~$\frakp_{L,\beta}(\BB)$ by further decomposing~$\BB$ into more elementary bad events.

Let~$\BBE$ denote the event that the block~$\Lambda_B$ contains an energetically ``charged'' site or bond. Explicitly,~$\BBE$ is the event that either for some~$\br\in\Lambda_B$ we have
\begin{equation}
\label{BBE}
|\Omega_{\br}\cdot\hate_y|\ge c_1\frac\Delta B,
\end{equation}
or, for some nearest-neighbor pair~$\langle\br,\br+\hate_\alpha\rangle$ in~$\Lambda_B$, we have
\begin{equation}
\bigl|\Omega_{\br}\cdot\hatv_{2\alpha}-\Omega_{\br+\hate_{\alpha}}\cdot\hatv_{2\alpha}\bigr|\ge c_2\frac\Delta B.
\end{equation}
Here~$c_1$ and~$c_2$ are constants that will be specified later.
The complementary part of~$\BB$ will be denoted by~$\BBSW$, i.e.,
\begin{equation}
\label{BBSW}
\BBSW=\BB\setminus\BBE.
\end{equation}
By the fact that~$\BBSW\subset\BBE^\cc$, on~$\BBSW$ the energetics of the entire block is good---i.e., the configuration is near one of the ground states. Clearly, all constant configurations with zero $y$-component are ground states. However, unlike for the 2D orbital-compass model, there are other, inhomogeneous ground states which make the treatment of this model somewhat more complicated. Fortunately, we will be able to plug in the results of~\cite{BCN1} more or less directly.

As for the orbital-compass model, to derive a good bound on~$\frakp_{L,\beta}(\BBSW)$ we will further partition~$\BBSW$ into more elementary events. We begin with the events corresponding to the homogeneous ground states: Given a collection of vectors~$\hatw_i$, $i=1,\dots,s$, that are uniformly spaced on the circle~$\scrS_1\subset\scrS_2$ in the~$xz$-plane, we define~$\BB_0^{(i)}$ to be the subset of~$\BBSW$ on which
\begin{equation}
\Omega_{\br}\cdot\hatw_i\ge\cos(\Delta),\qquad\br\in\Lambda_B.
\end{equation}
To describe the remaining ``parts of~$\BBSW$,'' we will not try to keep track of the entire ``near ground-state'' configuration. Instead, we will note that each inhomogeneous ground state contains a pair of neighboring planes in~$\Lambda_B$ where the homogenous configuration gets ``flipped'' through one of the vectors~$\hatv_1,\dots,\hatv_6$. 
(We refer the reader to \cite{BCN1}, particularly page~259.) Explicitly, given a lattice direction~$\alpha=1,2,3$ and a vector~$\hatw\in\scrS_1$, let~$\hatw_i^\star$ denote the reflection of~$\hatw_i$ through~$\hatv_{2\alpha-1}$. For each~$j=1,\dots,B-1$, we then define~$\BB_{\alpha,j}^{(i)}$ to be the set of spin configurations in~$\BBSW$ such that for all $\br\in\Lambda_B$,
\begin{equation}
\label{planar-states}
\begin{alignedat}{4}
\Omega_{\br}\cdot\hatw_i
&\ge\cos(\Delta)\qquad &\text{if}\qquad&\br\cdot\hate_\alpha=j,
\\
\Omega_{\br}\cdot\hatw_i^\star
&\ge\cos(\Delta)\qquad &\text{if}\qquad&\br\cdot\hate_\alpha=j+1.
\end{alignedat}
\end{equation}
(Note that~$\br\cdot\hate_\alpha=j$ means that the~$\alpha$-th coordinate of~$\br$ is~$j$. Hence, on~$\BB_{\alpha,j}^{(i)}$, the spins are near~$\hatw_i$ on the~$j$-th plane orthogonal 
to~$\hate_\alpha$ and near~$\hatw_i^\star$ on the~$j+1$-st plane in~$\Lambda_B$.)
The conditions under which these events form a partition of~$\BB$ is the subject of the following claim:

\begin{myproposition}
\label{prop-120-BBSW}
Given~$\kappa>0$, there exist~$c_1,c_2>0$ such that if $\BBE$ and $\BBSW$ are defined as in \twoeqref{BBE}{BBSW} and if $\Delta$ and~$B$ are such that $B\Delta\ll\kappa\ll1$ and~$s\Delta>4\pi$, then
\begin{equation}
\label{BBSW-decomp}
\BBSW\subseteq\bigcup_{i=1}^s\biggl(\,\BB_0^{(i)}\cup\bigcup_{\alpha=1,2,3}\bigcup_{j=1}^{B-1}\BB_{\alpha,j}^{(i)}\biggr)
\end{equation}
\end{myproposition}

Next we will attend to the estimates of~$\frakp_{L,\beta}$ for the various events constituting~$\BB$. As for the orbital-compass model, the event~$\BBE$ is dismissed easily:
\begin{equation}
\label{BBE-120}
\frakp_{L,\beta}(\BBE)\le c\beta B^3\texte^{-\tilde c\beta\Delta^2/B^2},
\end{equation}
where~$c$ and~$\tilde c$ are positive constants. As to the events~$\BB_0^{(i)}$, here we get:

\begin{myproposition}
\label{prop-120-homogeneous}
For each~$\kappa>0$ there exists~$\delta>0$ such that if $\beta$ and~$\Delta$ obey
\begin{equation}
\label{beta-Delta1}
\beta\Delta^2>\frac1\delta\quad\text{\rm and}\quad\beta\Delta^3<\delta,
\end{equation}
then for all~$L$ sufficiently large,
\begin{equation}
\frakp_{L,\beta}(\BB_0^{(i)})\le\texte^{-B^3\rho_1(\kappa)},
\qquad i=1,\dots,s.
\end{equation}
Here~$\rho_1(\kappa)>0$ for all~$\kappa\ll1$.
\end{myproposition}

For the ``inhomogeneous'' events the decay rate is slower, but still sufficient for our needs.

\begin{myproposition}
\label{prop-120-inhomogeneous}
For each~$\kappa>0$ there exists~$\delta>0$ such that if $\beta$, $\Delta$ and~$\delta$ obey \eqref{beta-Delta1}, then for all~$j=1,\dots,B-1$, all~$\alpha=1,2,3$ and all~$L$ sufficiently large,
\begin{equation}
\frakp_{L,\beta}(\BB_{\alpha,j}^{(i)})\le\texte^{-B^2\rho_2(\kappa)},
\qquad i=1,\dots,s.
\end{equation}
Here~$\rho_2(\kappa)>0$ for all~$\kappa\ll1$.
\end{myproposition}

Again, the proofs of these propositions are deferred to the Appendix.

\begin{proofsect}{Proof of Theorem~\ref{thm-120} completed}
We proceed very much like for the orbital compass model. The core of the proof again boils down to showing that~$\frakp_{L,\beta}(\BB)$ is small, provided~$B$ is chosen appropriately. Let~$\Delta$ and~$B$ be related to~$\beta$ as in \eqref{B-Delta}. By \eqref{BBE-120}, this choice makes~$\frakp_{L,\beta}(\BBE)$ small and, at the same time, makes \eqref{beta-Delta1} eventually satisfied for any fixed~$\delta>0$. Invoking Propositions~\ref{prop-120-homogeneous}-\ref{prop-120-inhomogeneous}, and the subadditivity of~$\AA\mapsto\frakp_{L,\beta}(\AA)$, we have
\begin{equation}
\frakp_{L,\beta}(\BBSW)\le s\bigl(\texte^{-B^3\rho_1(\kappa)}+3B\texte^{-B^2\rho_2(\kappa)}\bigr)
\end{equation}
which by the fact that~$s=O(\Delta^{-1})$ implies~$\frakp_{L,\beta}(\BBSW)\ll1$ once~$\beta$ is sufficiently large. Using that~$\frakp_{L,\beta}(\BB)\le\frakp_{L,\beta}(\BBE)+\frakp_{L,\beta}(\BBSW)$, the desired bound $\frakp_{L,\beta}(\BB)\ll1$ follows.

It is easy to check, the bad event~$\BB$ is preserved by ``complex conjugation''~$\sigma$ as well as reflections and so the~$\vartheta_{\bt}$'s act on it as mere translations. Moreover, once~$\kappa\ll1$, if two distinct types of goodness occur in neighboring blocks, all edges between the blocks are of high-energy---any block containing these edges is thus bad. Finally, the model on torus is invariant under rotation of all spins by~$60^\circ$ in the~$xz$-plane. This means that all conditions of Proposition~\ref{prop-SB} are satisfied and so, for~$\beta\gg1$ and~$\CalS\gg\beta^2$, the quantum model features six distinct states obeying \eqref{QQ-120}. From here we get \eqref{spin-120}.
\end{proofsect}


\section{Appendix}
\label{sec6}\noindent
This section is devoted to the proofs of various technical statements from Sects.~\ref{sec5.3},~\ref{sec5.4} and~\ref{sec5.5}. Some of the proofs in the latter two subsections are based on the corresponding claims from~\cite{BCN1,BCN2}. In such cases we will indicate only the necessary changes.

\subsection{Technical claims: Large-entropy models}
Consider a sequence~$(\frakc^{(p)})$ satisfying the large-entropy property and assume, without loss of generality, that $\Vert\frakc^{(p)}\Vert=1$ for all~$p\ge1$. Our goal here is to provide the bounds on~$\frakp_{L,\beta}(\BB)$ and the asymptotic statements concerning the dominance of the two types of goodness which were claimed in Propositions~\ref{prop-LE} and~\ref{prop5.10}. We begin with a lower estimate on the full partition function.

\begin{mylemma} 
\label{foist}
Let~$t>0$ be fixed. Then there exists~$p_1<\infty$ and constants~$c_1,c_2\in(0,\infty)$ such that for all~$p\ge p_1$ and all~$\beta\ge0$,
\begin{equation}
\liminf_{L\to\infty}\,(Z_L)^{1/L^d}\ge\max\bigl\{c_1\epsilon_p\,\texte^{\beta d A_p(t)},c_2\bigr\}.
\end{equation}
\end{mylemma}

\begin{proofsect}{Proof}
We will derive two separate bounds on the partition function per site. Focussing on the cases when~$\Omega_{\br}\diamond\Omega_{\br'}$ involves all three components of the spins, let us restrict attention to configurations when every spin is within angle~$c\sqrt{\epsilon_p}$ of the vector~$(0,0,1)$, where~$c$ is a constant to be determined momentarily. Let~$\Omega$ and~$\Omega'$ be two vectors with this property. Then the (diamond) angle between~$\Omega$ and~$\Omega'$ is less than~$2c\sqrt{\epsilon_p}$ and so
\begin{equation}
\Omega\diamond \Omega'\geq\cos\bigl(2c\sqrt{\epsilon_p}\bigr)\ge1-2c^2\epsilon_p.
\end{equation}
Choosing~$2c^2=t$, we thus have~$\Omega\diamond\Omega'\ge1-t\epsilon_p$. This means that the energy of any bond in the configuration obeying these constraint is at least~$A_p(t)$; while each spin has at least~$1-\cos(c\sqrt{\epsilon_p})\approx\frac12c^2\epsilon_p$ surface area at its disposal. This implies that~$(Z_L)^{1/L^d}$ is bounded by the first term in the maximum with~$c_1\approx\frac12c^2$. The other interpretation of~$\Omega_{\br}\diamond\Omega_{\br'}$ is handled analogously.

In order to derive the second bound, we will restrict all spins to a sector of angular aperture~$\ffrac\pi2$, e.g.,~the one described as $\{\Omega=(\Omega^1,\Omega^2,\Omega^3)\in \scrS^2\colon \Omega^1>1/\sqrt{2}\}$. This has area~$a$ which is a fixed positive number. Moreover, the constraint ensures that the interaction between any two spins is non-positive; the partition function per site then boils down to the entropy of such configurations. To evaluate this entropy, we fix the configuration on the even sublattice. Every spin on the even sublattice is then presented with $2d$ ``spots'' on this sector which it must avoid. The area of each such spot is a constant times~$\epsilon_p$. It follows that~$(Z_L)^{1/L^d}\ge a-O(\epsilon_p)$ which is positive once~$p$ is sufficiently large.
\end{proofsect}

Our next bound concerns the constrained partition function~$Z_L^{\mix}(\LL)$ obtained by disseminating a particular pattern~$\LL$ of ordered and disordered bonds (i.e.\ energetically and entropically good bonds) over the torus, when $\LL$ is a genuine mixture of the two. That is, we assume that $\LL$ contains bonds of both phenotypes. We remark that this dissemination is carried out by means of reflections in \emph{planes of sites} (which is permissible by the nearest-neighbor nature of the interaction).
Recall that $a_d=d2^{d-1}$ is the number of bonds entirely contained in the $2\times \cdots \times 2$ block $\Lambda_B$.

\begin{mylemma}
\label{lemma-Delta2}
Let $t>0$ be such that
\begin{equation}
\label{kappa2-cond}
\frac{1-(1-b)/a_d}{A_p(t)}\le1
\end{equation}
and
\begin{equation}
\label{Delta2}
\Delta\overset{\text{\rm def}}=\min\biggl\{1+\frac1{a_d}-\frac1{A_p(t)},\,
\frac1{a_d}-\frac{b}{A_p(t)}\biggr\}>0.
\end{equation}
Then there exists a constant~$c_3<\infty$ such that for any~$\beta\ge0$ and any pattern~$\LL$ of ordered and disordered bonds (i.e.\ energetically and entropically good bonds) on~$\Lambda_B$ containing at least one bond of each phenotype,
\begin{equation}
\label{6.4bd}
\limsup_{L\to\infty}\,Z^{\mix}_L(\LL)^{1/L^d}\le c_3\max\bigl\{c_1\epsilon_p\,\texte^{\beta d A_p(t)},c_2\bigr\}\,(\epsilon_p)^{\Delta}.
\end{equation}
\end{mylemma}

\begin{proofsect}{Proof}
Fix a pattern~$\LL$ as specified above. As usual, we call a bond disordered if it is entropically good. Let~$f_b$ denote the fraction of disordered bonds in pattern~$\LL$. Let us call a vertex  an ``entropic site" if all bonds connected to it are disordered.
(Note that this has two different, but logically consistent, connotations depending on whether we are speaking of a vertex in~$\Lambda_B$ or in~$\T_L$.)
Let~$f_s$ denote the fraction of entropic sites in~$\LL$.
Upon dissemination (by reflections through planes of sites), these numbers $f_b$ and $f_s$ will represent the actual fractions of disordered bonds and entropic sites in~$\T_L$, respectively. Now each disordered bond has an energetic at most~$b$, while we may estimate the energy of each ordered bond by~$1$. For each entropic site we will throw in full measure so we just need to estimate the entropy of the non-entropic sites. Here we note that each ordered bond disseminates into a ``line'' of ordered bonds, upon reflections. If we disregard exactly one bond on this ``line of sites'', then we see that there is a total measure proportional to $O(\epsilon_p^{L-1})$. Since this entropy is shared by the~$L$ vertices on this line, the entropy density of each vertex on this line is $O(\epsilon_p)$ in the $L\to\infty$ limit.
This is an upper bound for the entropy density for each non-entropic~site.

The bounds on energy show that the Boltzmann factor is no larger than
\begin{equation}
\texte^{\beta d(1-f_b)+\beta db f_b}=\texte^{\beta d[1-(1-b)f_b]}.
\end{equation}
We thus conclude that, for some constant~$\tilde c_3$,
\begin{equation}
\limsup_{L\to\infty}\,Z_L(\LL)^{1/L^d}\le
\tilde c_3(\epsilon_p)^{1-f_s}\texte^{\beta d[1-(1-b)f_b]}.
\end{equation}
Now, we may write the right-hand side as
\begin{equation}
\label{complicatedpower}
\tilde c_3\Bigl(\epsilon_p\texte^{\beta d A_p(t)}\Bigr)^{\textstyle\frac{1- (1-b) f_b}{A_p(t)}}
\,(\epsilon_p)^{\Delta(\LL)}
\end{equation}
where
\begin{equation}
\Delta(\LL)=1-f_s-\frac{1- (1-b) f_b}{A_p(t)}.
\end{equation}
Since $\LL$ contains at least one entropic bond, we know $f_b>1/a_d$.
Our choice of~$t$ guarantees that~$1- (1-b) f_b\le 1- (1-b)/a_d\le A_p(t)$ and so the complicated 
exponent in \eqref{complicatedpower} is bounded by 1. We may use the famous identity $X^{\lambda} Y^{1-\lambda} \leq \max(X,Y)$, true whenever $X,Y \geq 0$ and $0\leq \lambda\leq 1$,  to bound the term with the complicated power in~\eqref{complicatedpower} by the maximum in \eqref{6.4bd}. (We set $X=c_1\epsilon_p\,\texte^{\beta d A_p(t)}$ and $Y=c_2$, absorbing extra order-1 constants into our eventual $c_3$.) It remains to show that~$\Delta(\LL)$ exceeds~$\Delta$ in \eqref{Delta2} whenever~$\LL$ contains both phenotypes of bonds.

We will derive a relation between~$f_s$ and~$f_b$ that holds whenever~$\LL$ contains both phenotypes of bonds. We may give the argument in either picture---where we restrict to the small block $\Lambda_B$ or where we consider the full torus $\T_L$ after disseminating $\LL$---which are entirely equivalent since the fractions of entropic bonds and sites are the same. We will give the argument in the small $2\times \cdots \times 2$ block $\Lambda_B$. Since $\LL$ contains bonds of both phenotypes there are at least two vertices in $\Lambda_B$ each of which ``emanates" bonds of both phenotypes. We mark these sites, and for each of them we mark one of the incident entropically good (disordered) bonds. We now consider the bonds of $\Lambda_B$ to be split into half-bonds each of which is associated to the closest incident vertex (disregarding the midpoints). We label each half-bond as entropic or energetic, according to whether it is half of a full bond which is entropically or energetically good. 

Let $H$ be the total number of entropic half-bonds. Now note that for each entropic vertex, all $d$ of the half-bonds emanating from it (and contained in $\Lambda_B$) are ``entropic half-bonds". We also have at least two additional entropic half-bonds associated to the two marked sites. Therefore the number of entropic half bonds satisfies the bound $H \geq d 2^d f_s + 2$. (Note that there are $2^d f_s$ entropic sites.) Since there are $2a_d=d 2^d$ total half-bonds in $\Lambda_B$, the proportion of entropic half bonds is at least $f_s + 1/a_d$. At this point let us observe that the proportion of entropic half-bonds is exactly the same as the proportion of entropic full-bonds, $f_b$. Therefore
\begin{equation}
  f_b \geq f_s + \frac{1}{a_d}\, .
\end{equation}
Plugging this into the formula for~$\Delta(\LL)$ we thus get
\begin{equation}
\Delta(\LL)\ge1+\frac1{a_d}-f_b-\frac{1-(1-b)f_b}{A_p(t)}.
\end{equation}
Allowing~$f_b$ to take arbitrary values in~$[0,1]$, the right-hand side is minimized by one of the values in the maximum in~\eqref{Delta2}. Hence,~$\Delta(\LL)\ge\Delta$ whereby \eqref{6.4bd} follows.
\end{proofsect}

\begin{proofsect}{Proof of Proposition~\ref{prop-LE}}
As usual, we consider events disseminated by reflections in planes of lattice sites.
Let~$b_0<\frac1{1+a_d}$. If~$b\le b_0$, then, as a calculation shows, the bound \eqref{Delta2} holds as well as \eqref{kappa2-cond} for~$t$ such that~$A_p(t)\ge1-b$. Such a~$t$ can in turn be chosen by the assumption that the model obeys the large-entropy condition.
(This is where we need that~$p$ is sufficiently large.) Hence, the bound in Lemma~\ref{lemma-Delta2} is at our disposal. Now the maximum on the right-hand side of \eqref{6.4bd} is a lower bound on the full partition function per site; the lemma thus gives us bounds on~$\frakp_{L,\beta}$ of the events enforcing the various patterns on~$\Lambda_B$. Since~$\BB$ can be decomposed into a finite union of such pattern-events, the desired inequality \eqref{5.21two} follows. \end{proofsect}

\begin{proofsect}{Proof of Proposition~\ref{prop5.10}}
Again we work with events disseminated using reflections in planes of sites.
In order to prove \eqref{zero-bd}, we note that~$\frakE_p(\Omega_{\br}\diamond\Omega_{\br'})\ge b$---which is what every bond~$\langle\br,\br'\rangle$ in~$\Lambda_B$ satisfies provided~$\Omega\in\Gso$---implies $\Omega_{\br}\diamond\Omega_{\br'}\ge1-c\epsilon_p$. The neighboring spins are thus constrained to be within angle~$O(\sqrt{\epsilon_p})$ of each other. Disregarding an appropriate subset of these constraints (reusing the ``line of sites'' argument from the first part of the proof of Lemma \ref{lemma-Delta2}) the desired bound follows.

To prove \eqref{beta-zero-bd}, we note that the disseminated event~$\Gsd$ forces all bonds to have energy less than~$b$. Lemma~\ref{foist} implies that the corresponding~$\frakp_{L,\beta}$-functional is bounded above by $\tilde{C}_1 (\epsilon_p)^{-1}\texte^{\beta d[b-A_p(t)]}$. Assuming that $b<\ffrac{1}{2}$ and $t$ is chosen so that $A_p(t) -b > \ffrac{1}{2}$, we see that if~$\beta$ is large enough to satisfy
\begin{equation}
\texte^{\beta d} \geq \epsilon_p^{-2(1+\Delta_1)} ,
\end{equation}
then the $\frakp_{L,\beta}$~bound is less than~$\tilde{C}_1 (\epsilon_p)^{\Delta_1}$.
\end{proofsect}

Given the existing results on the discontinuity of energetic bonds, it is almost inconceivable that the
energy density itself could be continuous.
To mathematically rule out this possibility, we will show that, in actuality very few of the energetic bonds
have value in the vicinity of~$b$.
So while the previous argument only considered two types of bonds, we will
henceforth have the following three types of bonds:
\settowidth{\leftmargini}{(11)}
\begin{enumerate}
\item[(1)]
\emph{strongly ordered} if~$\frakE_p(\Omega_{\br}\diamond\Omega_{\br'})\ge1-b'$,
\item[(2)]
\emph{weakly ordered} if~$1-b'>\frakE_p(\Omega_{\br}\diamond\Omega_{\br'})\ge b$,
\item[(3)]
\emph{disordered} if~$\frakE_p(\Omega_{\br}\diamond\Omega_{\br'})<b$.
\end{enumerate}
Here $0<b',b<\ffrac{1}{2}$ are constants which we will choose later, although we already know that we have the restriction $b<1/(1+a_d)$ as was necessary in the proof of Proposition \ref{prop-LE}.
A rather similar line of argument to that previously used for mixed patterns of ordered and disordered bonds handles the situation for mixed patterns of weak and strong order.
For each pattern~$\LL$ of weakly and strongly ordered bonds on $\Lambda_B$, let~$Z_L^{\ord}(\LL)$ denote the partition function obtained by disseminating~$\LL$ all over the torus. Then we have:
\begin{mylemma}
\label{lemma-Delta1}
Let $t>0$ be a number such that
\begin{equation}
\label{Delta1}
\Delta' \overset{\text{\rm def}}=1-\frac{1-b'/a_d}{A_p(t)}>0.
\end{equation}
There exists a constant~$c_4<\infty$ such that for any $\beta\ge0$ and any pattern~$\LL$ of weakly and strongly ordered bonds on the~$2\times\cdots\times 2$ block~$\Lambda_B$ containing at least one weakly ordered bond,
\begin{equation}
\label{6.5bd}
\limsup_{L\to\infty}\,Z_L^{\ord}(\LL)^{1/L^d}\le c_4\,\max\bigl\{c_1\epsilon_p\,\texte^{\beta d A_p(t)},c_2\bigr\}\,(\epsilon_p)^{\Delta'}.
\end{equation}
\end{mylemma}

\begin{proofsect}{Proof}
Consider an ordered pattern~$\LL$ with fraction~$f_w$ of weakly ordered bonds. After dissemination all over~$\T_L$, there is a  fraction~$f_w$ of bonds on~$\T_L$ that are weakly ordered and a fraction fraction~$1-f_w$ that are strongly ordered. Putting energy~$1-b'$ for each weakly ordered bond and~$1$ for each strongly ordered bond, the Boltzmann weight of any spin configuration contributing to~$Z_L^\ord(\LL)$ is at most
\begin{equation}
\texte^{\beta d(1-b')f_w+\beta d(1-f_w)}=\texte^{\beta d(1-b' f_w)}.
\end{equation}
To calculate the entropy, we again use the ``line of sites'' argument from the first part of the proof of Lemma \ref{lemma-Delta2}, which gives an entropy per site on the order of $O(\epsilon_p)$ in the $L\to \infty$ limit.
This implies that the limsup of~$Z_L^\ord(\LL)^{1/L^d}$ is bounded by a constant times~$\epsilon_p\texte^{\beta d(1-b' f_w)}$. Since~$1-b' f_w \le1-b'/a_d$ we get
\begin{equation}
\limsup_{L\to\infty}\,Z_L^{\ord}(\LL)^{1/L^d}\le\tilde c_4
\Bigl(\epsilon_p\texte^{\beta d A_p(t)}\Bigr)^{\textstyle\frac{1-b'/a_d}{A_p(t)}}
(\epsilon_p)^{\Delta'},
\end{equation}
for some constant~$\tilde c_4<\infty$.
By \eqref{Delta1}, the exponent of the term~$\epsilon_p\texte^{\beta d A_p(t)}$ is less than~$1$ and so the famous identity, $X^\lambda Y^{1-\lambda}\le\max\{X,Y\}$, may be used again (as in the proof of Lemma \ref{lemma-Delta2}) which readily yields the bound \eqref{6.5bd}.
\end{proofsect}

\begin{proofsect}{Proof of Corollary  \ref{cor-Discont}}
The proof is based on thermodynamical arguments. First, standard calculations using coherent states show that
\begin{equation}
\frakE_p\bigl(\CalS^{-2}(S_{\br}\diamond S_{\br'})\bigr)\ket\Omega
=\frakE_p(\Omega_{\br}\diamond\Omega_{\br'}) \ket{\Omega}+O(1/\sqrt{\CalS})
\end{equation}
where the error term depends implicitly on~$p$. Hence, for a given~$p$ and~$\delta>0$, we can find~$\CalS$ so large that for any~$\br,\br'\in\Lambda_B$
\begin{equation}
\frac{
\bigl\langle\Omega\big|\frakE_p\bigl(\CalS^{-2}(S_{\br}\diamond S_{\br'})\bigr)
\hatQ_{\AA}\big|\Omega\bigr\rangle}
{\bigl\langle\Omega\big|\hatQ_{\AA}\big|\Omega\bigr\rangle}
\begin{cases}
\ge1-b'-\delta,\quad&\text{if }\AA=\Gso,
\\
\le b+\delta,\quad&\text{if }\AA=\Gsd.
\end{cases}
\end{equation}
(At the classical level the second case is by definition, whereas the first case follows from Lemma~\ref{lemma-Delta1}.)
Since~$\beta\mapsto\frake(\beta)$ is increasing, we conclude that \eqref{e-beta-jump} holds.
As a technical point, we note that in the statement of the corollary we did not include the small corrections corresponding to $\delta>0$. This was primarily for \ae{}sthetic reasons: we wanted to state the simplest possible result. We can clearly accomplish this by taking~$b$ and~$b'$ to be a little smaller than is otherwise needed.
\end{proofsect}

\subsection{Technical claims: Orbital-compass model}
Here we will prove Propositions~\ref{prop-SWbd}-\ref{prop-FEmin} concerning the orbital-compass model. The proofs follow the strategy developed in the context of the 120-degree model~\cite{BCN1}.

\begin{proofsect}{Proof of Proposition~\ref{prop-SWbd}}
The proof goes by one more partitioning of~$\BBSW^{(i)}$. Consider a spin configuration~$\Omega=(\Omega_{\br})_{\br\in\T_L}\in\BBSW^{(i)}$. Since~$\BBSW^{(i)}\subset\BBSW$ and~$\Delta\ll1$, it is easy to check the following facts:
\settowidth{\leftmargini}{(11)}
\begin{enumerate}
\item[(1)]
the~$y$-components of all spins in~$\Lambda_B$ are small.
\item[(2)]
the $x$-components of the spins along each ``line of sites'' (in~$\Lambda_B$) in the $x$-direction are either all near the $x$-component of vector~$\hatw_i$ or its negative.
\item[(3)]
same is true for the~$z$-components of the spins on ``lines of sites'' in the~$z$ lattice direction.
\end{enumerate}
Thus, at the cost of reflecting the $x$-components of spins along each ``line of sites'' in the $x$-direction, and similarly for the $z$-components, we may assume that all spins are aligned with~$\hatw_i$ in the sense that
\begin{equation}
\label{BBSWi0}
\Omega_{\br}\cdot\hatw_i\ge\cos(\Delta),\qquad \br\in\Lambda_B.
\end{equation}
Let~$\BBSW^{(i,0)}$ denote the set of configurations satisfying \eqref{BBSWi0}.
The above reflection preserves both the \emph{a priori} measure and the Hamiltonian \eqref{OCMclass};
the event $\BBSW^{(i)}$ is thus partitioned into $2^{2B}$ ``versions'' of event~$\BBSW^{(i,0)}$ all of which have the same value of~$\frakp_{L,\beta}$-functional. Invoking the Subadditivity Lemma, \eqref{SWbd} is proved once we show that
\begin{equation}
\frakp_{L,\beta}(\BBSW^{(i,0)})\le \texte^{-B^2(F_{L,\Delta}(\hatw_i)-F_{L,\Delta}(\hate_1))}.
\end{equation}
This follows by noting that $\texte^{-B^2\,F_{L,\Delta}(\hatw_i)}$ is, to within a convenient multiplier, the integral of  the Boltzmann weight $\texte^{-\beta H^\infty(\Omega)}$ on the event~$\BBSW^{(i,0)}$ while $\texte^{-B^2F_{L,\Delta}(\hate_1)}$ provides a lower bound on the partition function (again, to within the same multiplier which thus cancels from the ratio).
\end{proofsect}

\begin{proofsect}{Proof of Proposition~\ref{prop-FElim}}
The principal idea is to derive upper and lower bounds on $F_{L,\Delta}(\hatw)$ which converge, in the limit~$L\to\infty$, to the same Gaussian integral. Let us parametrize~$\hatw\in\scrS_1^{++}$ as~$(\cos\theta_\star,0,\sin\theta_\star)$ and, given a spin configuration~$\Omega$ that satisfies $\Omega_{\br}\cdot\hatw\ge\cos(\Delta)$ for all~$\br\in\T_L$, let us introduce the deviation variables~$(\vartheta_{\br},\zeta_{\br})$ by the formula
\begin{equation}
\label{Omega-zeta}
\Omega_{\br}=\Bigl(\sqrt{1-\zeta_{\br}^2}\cos(\theta_\star+\vartheta_{\br}),\zeta_{\br},\sqrt{1-\zeta_{\br}^2}\,\sin(\theta_\star+\vartheta_{\br})\Bigr).
\end{equation}
Noting that both~$\vartheta_{\br}$ and~$\zeta_{\br}$ are order~$\Delta$, we derive that $H^\infty(\Omega)+|\T_L|$ is, to within a quantity of order~$L^2\Delta^3$, equal to the quadratic form
\begin{equation}
\scrI_{L,\hatw}(\vartheta,\zeta)=\frac12\sum_{\br\in\T_L}\Bigl\{\hatw_z^2\,(\vartheta_{\br}-\vartheta_{\br+\hate_x})^2+\hatw_x^2\,(\vartheta_{\br}-\vartheta_{\br+\hate_z})^2\Bigr\}+\sum_{\br\in\T_L}\zeta_{\br}^2
\end{equation}
The Jacobian of the transformation $\Omega_{\br}\mapsto(\vartheta_{\br},\zeta_{\br})$ is unity.

Next we will derive upper and lower bounds on the integral of~$\texte^{-\beta\scrI_{L,\hatw}}$ against the product of indicators in \eqref{FLDelta}. For the upper bound we invoke the inequality
\begin{equation}
\prod_{\br\in\T_L}\1_{\{\Omega_{\br}\cdot\hatw\ge\cos(\Delta)\}}
\le\texte^{\frac12\lambda\beta L^2\Delta^2}\exp\Bigl\{-\frac{\lambda\beta}2\sum_{\br\in\T_L}\vartheta_{\br}^2\Bigr\},
\end{equation}
valid for each~$\lambda\ge0$.
The~$\zeta_{\br}$'s are then unrestricted and their integrals can be performed yielding a factor~$\sqrt{2\pi/\beta}$ per integral. The integral over~$\vartheta_{\br}$'s involves passing to the Fourier components, which diagonalizes the covariance matrix. The result is best expressed in~$L\to\infty$ limit:
\begin{equation}
\label{OCM-UB}
\liminf_{L\to\infty}F_{L,\Delta}(\hatw)\ge O(\beta\Delta^3)+\frac12\lambda\beta\Delta^2+F(\lambda,\hatw),
\end{equation}
where
\begin{equation}
F(\lambda,\hatw)=\frac12\int_{[-\pi,\pi]^2}\frac{\textd\bk}{(2\pi)^2}\log\bigl[\lambda+\wh D_{\bk}(\hatw)\bigr]
\end{equation}
By the Monotone Convergence Theorem,~$F(\lambda,\hatw)$ converges to~$F(\hatw)$ as~$\lambda\downarrow0$. Since $\beta\Delta^3$ is less than~$\delta$, which is up to us to choose, taking~$\lambda\downarrow0$ on both sides of \eqref{OCM-UB} we deduce that $F_{L,\Delta}(\hatw)\ge F(\hatw)-\epsilon$ for~$L$ sufficiently large. 

It remains to derive the corresponding lower bound. Here we will still work with the parameter~$\lambda$ above but, unlike for the upper bound, we will not be able to take~$\lambda\downarrow0$ at the end. Consider the Gaussian measure~$P_\lambda$ which assigns any Borel set~$\AA\subset(\R\times\R)^{\T_L}$ the probability
\begin{equation}
P_\lambda(\AA)=\frac1{Z_L(\lambda)}\int_{\AA}
\Bigl(\frac\beta{2\pi}\Bigr)^{\T_L}
\exp\Bigl\{-\beta\scrI_{L,\hatw}(\vartheta,\zeta)-\frac{\beta\lambda}2\sum_{\br\in\T_L}\vartheta_{\br}^2\Bigr\}
\prod_{\br\in\T_L}\textd\vartheta_{\br}\textd\zeta_{\br}.
\end{equation}
Let~$E_\lambda$ denote the corresponding expectation. From~$\beta\lambda\ge0$ we get
\begin{equation}
\int_{(\scrS_2)^{|\T_L|}}\!\!\textd\Omega\,\,\texte^{-\beta\scrI_{L,\hatw}(\vartheta,\zeta)}\,\biggl(\,\prod_{\br\in\T_L}\1_{\{\Omega_{\br}\cdot\hatw\ge\cos(\Delta)\}}\biggr)
\ge Z_L(\lambda)\,E_\lambda\biggl(\,\prod_{\br\in\T_L}\1_{\{\Omega_{\br}\cdot\hatw\ge\cos(\Delta)\}}\biggr).
\end{equation}
The free-energy corresponding to the normalization constant~$Z_L(\lambda)$ is exactly~$F(\lambda,\hatw)$ above. Thus, given~$\epsilon>0$, we can find~$\lambda>0$ such that~$Z_L(\lambda)\ge\texte^{-L^2[F(\hatw)+\epsilon/2]}$ once~$L\gg1$. It remains to show that the expectation is at least~$\texte^{-L^2\epsilon/2}$ provided~$\delta$ in \eqref{beta-Delta} is sufficiently small.

Here we first decrease the product by noting that
\begin{equation}
\1_{\{\Omega_{\br}\cdot\hatw\ge\cos(\Delta)\}}
\ge
\1_{\{|\vartheta_{\br}|\le\ffrac\Delta2\}}
\1_{\{|\zeta_{\br}|\le\ffrac\Delta2\}}.
\end{equation}
This decouples the~$\zeta_{\br}$'s from the~$\vartheta_{\br}$'s and allows us to use the independence of these fields under~$P_\lambda$. Since the~$\zeta_{\br}$'s are themselves independent, the integral over~$\zeta_{\br}$ boils down to
\begin{equation}
E_\lambda\biggl(\,\prod_{\br\in\T_L}\1_{\{|\zeta_{\br}|\le\ffrac\Delta2\}}\biggr)
=\prod_{\br\in\T_L}\,P_\lambda\bigl(|\zeta_{\br}|\le\ffrac\Delta2\bigr)
\ge\bigl(1-\texte^{-\lambda\beta\Delta^2/4}\bigr)^{L^2},
\end{equation}
where we used the standard tail bound for normal distribution.
Note that, for any fixed $\lambda>0$, the term $1-\texte^{-\lambda\beta\Delta^2/4}$ can be made as close to one as desired by increasing~$\beta\Delta^2$ appropriately.

The~$\vartheta_{\br}$'s are not independent, but reflection positivity through bonds shows that the corresponding indicators are positively correlated, i.e.,
\begin{equation}
E_\lambda\biggl(\,\prod_{\br\in\T_L}\1_{\{|\vartheta_{\br}|\le\ffrac\Delta2\}}\biggr)
\ge \prod_{\br\in\T_L}\,P_\lambda\bigl(|\vartheta_{\br}|\le\ffrac\Delta2\bigr).
\end{equation}
The probability on the right-hand side is estimated using a variance bound:
\begin{equation}
P_\lambda\bigl(|\vartheta_{\br}|>\ffrac\Delta2\bigr)
\le\Bigl(\frac2\Delta\Bigr)^2\text{\rm Var}(\vartheta_{\br})
=\frac4{\Delta^2}\frac1{L^2}\sum_{\bk\in\T_L^\star}\frac1{\beta[\lambda+\wh D_{\bk}(\hatw)]}
\le\frac4{\lambda\beta\Delta^2},
\end{equation}
where~$\T_L^\star$ denotes the reciprocal torus. Again, for any fixed~$\lambda$, $P_\lambda(|\vartheta_{\br}|\le\ffrac\Delta2)$ can be made as close to one as desired once~$\beta\Delta^2$ is sufficiently large. We conclude that, given~$\epsilon>0$, we can choose~$\delta$ such that $F_{L,\Delta}(\hatw)\le F(\hatw)+\epsilon$ once~$L\gg1$. This finishes the proof.
\end{proofsect}

\begin{proofsect}{Proof of Proposition~\ref{prop-FEmin}}
Since~$\hatw_x^2+\hatw_z^2=1$, this is a simple consequence of Jensen's inequality and the strict concavity of the logarithm.
\end{proofsect}

\subsection{Technical claims: 120-degree model}
\noindent
Here we will provide the proofs of technical Propositions~\ref{prop-120-BBSW}-\ref{prop-120-inhomogeneous}. The core of all proofs is the fact that any spin configuration $(\Omega_{\br})$ can be naturally deformed, by rotating along the main circle orthogonal to the~$xz$-plane, to have zero~$y$-component. An explicit form of this transformation is as follows: Let us write each $\Omega_{\br}\in\scrS_2$ using two variables~$\zeta_{\br}\in[-1,1]$ and~$\theta_{\br}\in[0,2\pi)$ interpreted as the cylindrical coordinates,
\begin{equation}
\Omega_{\br}=\Bigl(\,\sqrt{1-\zeta_{\br}^2}\,\cos\theta_{\br},\,\zeta_{\br},\,\sqrt{1-\zeta_{\br}^2}\,\sin\theta_{\br}\Bigr).
\end{equation}
Then~$\Omega_{\br}'$ is the vector in which we set~$\zeta_{\br}=0$, i.e.,
\begin{equation}
\Omega_{\br}'=(\cos\theta_{\br},0,\sin\theta_{\br}).
\end{equation}
(We have already used this transformation in the proof of Proposition~\ref{prop-FElim}.) An additional useful feature of this parametrization is that the surface (Haar) measure~$\textd\Omega_{\br}$ on~$\scrS_2$ then decomposes into the product of the Lebesgue measure~$\textd\Omega_{\br}'$ on~$\scrS_1$ and the Lebesgue measure~$\textd\zeta_{\br}$ on~$[-1,1]$. 

\begin{proofsect}{Proof of Proposition~\ref{prop-120-BBSW}}
We will use the fact that, for configurations on~$\Lambda_B$ with \emph{vanishing} component in the~$y$-direction, this was already proved as Theorem~6.4 in~\cite{BCN1}. Let~$(\Omega_{\br})\in\BBSW$ and define~$(\Omega_{\br}')$ be as above. Since~$|\Omega_{\br}\cdot\hate_y|\le c_1\Delta/B$ for all~$\br\in\Lambda_B$, we have
\begin{equation}
\bigl|(\Omega_{\br}-\Omega_{\br}')\cdot\hate_y\bigr|\le c_1\Delta/B
\end{equation}
while
\begin{equation}
(\Omega_{\br}-\Omega_{\br}')\cdot\hate_\alpha=O(\Delta^2/B^2),
\qquad\alpha=x,z.
\end{equation}
In particular, the configuration $(\Omega_{\br}')$ is contained in the version of event~$\BBSW$ from~\cite{BCN1}, provided~$c_2$ is a sufficiently small numerical constant. Thus, under the condition $B\Delta\ll\kappa\ll1$---which translates to the condition~$B\sqrt\Gamma\ll\kappa\ll1$ of~\cite[Theorem~6.4]{BCN1}---$(\Omega_{\br}')$ is contained in one of the events on the right-hand side of \eqref{BBSW-decomp}. But, at the cost of a slight adjustment of~$\Delta$, the corresponding event will then contain also~$(\Omega_{\br})$.
\end{proofsect}

To prove the bounds in the remaining two propositions, we will more or less directly plug in the results of \cite{BCN1}. This is possible because the $y$-component of the spins contributes only an additive factor to the overall spin-wave free energy. The crucial estimate is derived as follows:

\begin{mylemma}
\label{lemma-reduce}
There exists a constant~$c>0$ such that the following is true: Let $\Delta\ll1$ and let~$\Omega=(\Omega_{\br})$ be a configuration on~$\T_L$ such that $|\Omega_{\br}\cdot\hate_y|\le\Delta^2$ and~$|\Omega_{\br}^{(2\alpha)}-\Omega_{\br+\hate_\alpha}^{(2\alpha)}|\le\Delta$, for all~$\alpha=1,2,3$. Define~$\Omega'=(\Omega_{\br}')$ as above. Then
\begin{equation}
\biggl|\,H^\infty(\Omega)-H^\infty(\Omega')-\frac32\sum_{\br\in\T_L}(\Omega_y\cdot\hate_y)^2\biggr|\le c\Delta^3 L^3.
\end{equation}
\end{mylemma}

\begin{proofsect}{Proof}
By the fact that~$\Omega_{\br}\cdot\hate_y=O(\Delta)$ we have
\begin{equation}
\Omega_{\br}\cdot\hatv_\alpha=\Omega_{\br}'\cdot\hatv_\alpha+O(\Delta^2).
\end{equation}
But then the assumption $\Omega_{\br}^{(2\alpha)}-\Omega_{\br+\hate_\alpha}^{(2\alpha)}=O(\Delta)$ yields
\begin{equation}
\bigl[(\Omega_{\br}-\Omega_{\br+\hate_\alpha})\cdot\hatv_{2\alpha}\bigr]^2
=\bigl[(\Omega_{\br}'-\Omega_{\br+\hate_\alpha}')\cdot\hatv_{2\alpha}\bigr]^2+O(\Delta^3).
\end{equation}
Using \eqref{120-class}, this proves the claim.
\end{proofsect}

\begin{proofsect}{Proof of Proposition~\ref{prop-120-homogeneous}}
The quantity~$\frakp_{L,\beta}(\BB_0^{(i)})$ is the ratio of the partition function in which all spins are constrained to make angle at most~$\Delta$ with~$\hatw_i$, and the full partition function. The restriction $\BB_0^{(i)}\subset\BBSW$ can, for the most part, be ignored except for the~$\hatw_i$'s that are close to one of the six preferred directions. In such cases the fact that~$\Delta\ll\kappa$ tells us that~$\BB_0^{(i)}$ is empty whenever the angle between~$\hatw_i$ and the closest of~$\hatv_1,\dots,\hatv_6$ is less than, say,~$\ffrac\kappa2$. In particular, we may restrict attention to the~$\hatw_i$'s that are farther than~$\ffrac\kappa2$ from any of these vectors.

Viewing the collection of angles~$(\theta_{\br})$ as a configuration of~$O(2)$-spins, Lemma~\ref{lemma-reduce} tells us that the Hamiltonian of~$(\Omega_{\br})$ is, to within corrections of order~$L^3\Delta^3$, the sum of $\frac32\sum_{\br}\zeta_{\br}^2$ and the Hamiltonian of the classical, $O(2)$-spin 120-degree model evaluated at configuration~$(\theta_{\br})$. Since the measure~$\textd\Omega_{\br}$ equals the product~$\textd\zeta_{\br}\textd\theta_{\br}$ on the respective domain, we may ignore the restriction of~$\zeta_{\br}$ to values less than~$O(\Delta)$ and integrate the~$\zeta_{\br}$'s. We conclude that~$\frakp_{L,\beta}(\BB_0^{(i)})$ is bounded by the same quantity as for the $O(2)$-spin 120-degree model times~$\texte^{O(\beta\Delta^3)}$. Since~$\beta\Delta^3$ is controlled via \eqref{beta-Delta1}, the desired bound follows from \cite[Lemma~6.9]{BCN1}.
\end{proofsect}

\begin{proofsect}{Proof of Proposition~\ref{prop-120-inhomogeneous}}
The proof is very much like that of the previous proposition. Let~$\wt\BB_{\alpha,j}^{(i)}$ denote the event that the top line in \eqref{planar-states} holds for all~$\br\in\Lambda_B$ for which~$\br\cdot\hate_\alpha$ is odd and the bottom line for all such~$\br$ for which~$\br\cdot\hate_\alpha$ is even. Chessboard estimates then yield
\begin{equation}
\frakp_{L,\beta}\bigl(\BB_{\alpha,j}^{(i)}\bigr)\le\frakp_{L,\beta}\bigl(\wt\BB_{\alpha,j}^{(i)}\bigr)^{2/B}.
\end{equation}
On the disseminated event~$\bigcap_{\bt\in\T_{L/B}}\theta_{\bt}(\wt\BB_{\alpha,j}^{(i)})$ the assumptions of Lemma~\ref{lemma-reduce} are satisfied. Hence, we may again integrate out the~$\zeta_{\br}$'s to reduce the calculation to that for $O(2)$-spins. The latter calculation was performed in detail in~\cite{BCN1}; the desired bound is then proved exactly as Lemma~6.10 of~\cite{BCN1} (explicitly, applying inequality (6.24) of~\cite{BCN1} and the paragraph thereafter).
\end{proofsect}

\section*{Acknowledgments}
\noindent 
This research was supported by the NSF grants~DMS-0306167 and~DMS-0505356. The authors wish to thank Elliott Lieb, Aernout van Enter and the anonymous referees for may useful comments on the first version of this paper.

\end{document}